\documentclass[a4paper,fleqn]{cas-dc}

\usepackage[numbers, sort&compress]{natbib}
\usepackage{graphicx}
\usepackage{hyperref}
\usepackage[hyphenbreaks]{breakurl}
\usepackage{psfrag}
\usepackage{bm}
\usepackage{amsmath}
\usepackage{amssymb}

\usepackage{caption}
\usepackage{multirow}
\usepackage{diagbox}
\usepackage{threeparttable}
\usepackage{epstopdf}
\usepackage{lscape}

\def\tsc#1{\csdef{#1}{\textsc{\lowercase{#1}}\xspace}}
\tsc{WGM}
\tsc{QE}

\begin{document}
\begin{sloppypar}
\let\WriteBookmarks\relax
\def\floatpagepagefraction{1}
\def\textpagefraction{.001}

\shorttitle{Short-length SSVEP data extension by a novel generative adversarial networks based framework}

\shortauthors{Y. Pan et al.}

\title [mode = title]{Short-length SSVEP data extension by a novel generative adversarial networks based framework}

\author[1]{Yudong Pan}
\credit{Methodology, Software, Validation, Writing-Original Draft, 
	Writing-Review and Editing, Visualization}	

\author[1]{Ning Li}
\credit{Methodology, Writing-Review}	

\author[1,2,3]{Yangsong Zhang\corref{cor1}}
\credit{Supervision, Conceptualization, Methodology, 
	Writing-Review and Editing}
\ead{zhangysacademy@gmail.com}

\author[3]{Peng Xu}
\credit{Writing-Review and Editing}

\author[3]{Dezhong Yao}
\credit{Writing-Review and Editing}

\cormark[1]




\affiliation[1]{organization={School of Computer Science and Technology, Laboratory for Brain Science and Medical Artificial Intelligence, Southwest University of Science and Technology},
            city={Mianyang},
            postcode={621010},
            country={China}}

\affiliation[2]{organization={Key Laboratory of Testing Technology for Manufacturing Process, Ministry of Education, Southwest University of Science and Technology},
				city={Mianyang},
			    postcode={621010},
		        country={China}}

\affiliation[3]{organization={MOE Key Lab for Neuroinformation, University of Electronic Science and Technology of China},
            city={Chengdu},
            postcode={610059},
            country={China}}

\cortext[1]{Corresponding author}



\begin{abstract}
	Steady-state visual evoked potentials (SSVEPs) based brain-computer interface (BCI) has received considerable attention due to its high information transfer rate (ITR) and available quantity of targets. However, the performance of frequency identification methods heavily hinges on the amount of user calibration data and data length, which hinders the deployment in real-world applications. Recently, generative adversarial networks (GANs)-based data generation methods have been widely adopted to create synthetic electroencephalography (EEG) data, holds promise to address these issues. In this paper, we proposed a GAN-based end-to-end signal transformation network for Time-window length Extension, termed as TEGAN. TEGAN transforms short-length SSVEP signals into long-length artificial SSVEP signals. By incorporating a novel U-Net generator architecture and an auxiliary classifier into the network architecture, the TEGAN could produce conditioned features in the synthetic data. Additionally, we introduced a two-stage training strategy and the LeCam-divergence regularization term to regularize the training process of GAN during the network implementation. The proposed TEGAN was evaluated on two public SSVEP datasets (a 4-class dataset and a 12-class dataset). With the assistance of TEGAN, the performance of traditional frequency recognition methods and deep learning-based methods have been significantly improved under limited calibration data. And the classification performance gap of various frequency recognition methods has been narrowed. This study substantiates the feasibility of the proposed method to extend the data length for short-time SSVEP signals for developing a high-performance BCI system. The proposed GAN-based methods have the great potential of shortening the calibration time and cutting down the budget for various real-world BCI-based applications.  
\end{abstract}




\maketitle

\section{Introduction}\label{sec1}
Brain-computer interface (BCI) has shown to become a promising technology that can provide its users with communication channels by decoding their neural activities into specific control commands, which do not depend on conventional output channels of peripheral nerves and muscles \cite{wolpaw2000BCI}. Among various neuroimaging modalities to implement a BCI system, electroencephalography (EEG) has been the most prominent one accounting for such the advantages as non-invasiveness, high temporal resolution, affordability, ease of implementation, portability, and convenience of use \cite{hwang2013EEG, abiri2019EEG}. 

There are several most popular paradigms can be employed to build EEG-based BCI systems, such as motor imaginary (MI) \cite{ang2008FBCSP}, P300 event related potentials (P300) \cite{nijboer2008P300}, auditory steady-state response (ASSR) \cite{kim2011ASSR}, and steady-state visual evoked potential (SSVEP) \cite{zhang2012MFSC}. Among them, SSVEP-based BCI systems have received considerable attention due to its high information transfer rate (ITR) and available quantity of targets. SSVEPs refer to periodic evoked potentials over occipital scalp areas, in response to rapidly repetitive visual stimulation flicking or reversing at a specific frequency \cite{lim2013SSVEP}. The SSVEP signal consists of a number of discrete frequency components, normally including the fundamental frequency of the visual stimulus and its harmonics. On the strengths and characteristics of SSVEP, numerous SSVEP-based BCI applications have been developed, such as bionic mechanical leg \cite{kwak2017CNN}, unmanned aerial vehicle \cite{dang2021MHLCNN}, dial interface \cite{nakanishi2015eCCA}, high-speed mental speller \cite{chen2015Speller}, smart homes \cite{kim2019SmartHome}, and games \cite{lalor2005GAME}. To design a high-performance BCI system based on SSVEP paradigms, the most crucial aspect is to develop a fast and accurate frequency recognition method that can distinguish the stimulus frequency of the target gazed by the users through analyzing the EEG signal in the shortest possible time. Thus, various cutting-edge algorithms have been proposed based on different perspectives \cite{Pan2023Survey}.

Since the SSVEP frequency is the same as the stimulus frequency, a typical algorithm used in early stage was the power spectral density analysis (PSDA) based on fast Fourier transform (FFT) \cite{hakvoort2011PSDA}. This method works on a single channel, thereby an optimal channel selection operation is required. Without troublesome optimization procedure, training-free multi-channel frequency recognition methods such as canonical correlation analysis (CCA) \cite{lin2006CCA} and multivariate synchronization index (MSI) \cite{zhang2014MSI} based on sine-cosine reference signals become prevalent for a long period of time. CCA and MSI are able to achieve comparable performance when only a few stimulus targets are available and the EEG signals are sufficiently long, but their performance drops dramatically when encountering a large number of targets and short-time signals \cite{chen2021implementing}. One possible explanation for this phenomenon is that more target stimuli and shorter signals make subject-independent sine-cosine template signals more susceptible to spontaneous EEG interference. To overcome this problem, individual template CCA (ITCCA) \cite{bin2011ITCCA} incorporates the subject-dependent training data to calculate the template signal and have made significant performance improvement under these hard conditions. The success of using individual training data opens a new era for the development of frequency recognition algorithms. Since then, many supervised spatial filtering algorithms have been proposed and become dominant in 40-class SSVEP-BCI systems, such as task-related component analysis (TRCA) \cite{nakanishi2017TRCA}, and correlated component analysis (CORCA) \cite{zhang2018CORCA}, etc. These methods need abundant individual training data i.e., user calibration data to calculate the spatial filters for all the stimulus frequencies. However, collecting user calibration data is a time-consuming and laborious process, prolonged experiments would even cause user’s fatigue, leading to the decreased quality of SSVEP signals. Hence, how to develop a high-performance frequency recognition algorithm that is calibration-free or needs only a small amount of calibration data has become a hot research topic in recent years \cite{chiang2021LST, wong2020stCCA, wong2021tlCCA, yan2022CSSFT}.

Benefit from the rapid development of deep learning (DL) in the past decade, there has been an increased interest in applying DL algorithms to detect SSVEPs \cite{Pan2023Survey}. In view of powerful feature representation capacity and flexibility, the DL-based methods hold promise for reducing the dependency of user calibration data. A significant advantage of DL methods is that they can extract common features and connections between similar data from massive amounts of data. Therefore, a feasible route is transferring the knowledge of existing subject database to the classifier of unseen subject. For instance, Waytowich et al. employed a compact convolutional neural network (Compact-CNN) entitled EEGNet to process raw time domain data and conduct inter-subject classification, which yielded about 80\% accuracy in a 12-class SSVEP dataset without any user calibration data \cite{waytowich2018EEGNet}. Ravi et al. utilized fast Fourier transform (FFT) to convert raw time domain data into frequency domain data and design a complex spectrum CNN (C-CNN) which could be trained on user-independent scheme, achieving about 81.6\%  accuracy on a 12-class and a 7-class SSVEP dataset, respectively \cite{ravi2020CCNN}. Chen et al. introduced a Transformer-based deep neural network model named SSVEPformer for enhancing the performance of zero-calibration SSVEP-BCI \cite{chen2023SSVEPformer}. The experimental results have shown that SSVEPformer could achieve high accuracy of about 84\% and 80\% on a 12-class and 40-class SSVEP dataset. Although the inter-subject classification conducted in these studies that do not require the usage of target subject calibration data, the performance still has large room for improvement compared to the intra-subject classification. Considering that another prominent characteristic of DL methods is that they can handle complex multivariate problems by learning a large number of neuron parameters to fit any nonlinear function and combine regularization techniques to reduce the demand for training data. Our previous study proposed an efficient CNN-LSTM (Long short-term memory) network with spectral normalization and label smoothing technologies, termed as SSVEPNet, for intra-subject classification under the small sample size and short-time window scenarios \cite{pan2022SSVEPNet}. SSVEPNet was verified on a 4-class and a 12-class SSVEP dataset, yielding about 88.0\% accuracy when a few trials of each stimulus available. 

On the other hand, in addition to explicitly modelling an efficient frequency recognition method that requires less calibration data, recent studies have substantiated the potential of using the generative models to address the issues of data shortage in the SSVEP classification tasks \cite{luo2022SAME}. Generative models aim to learn the distribution of real data by constructing a probabilistic statistical model given real data and using it to generate synthetic data that approximate the distribution of real data \cite{thirumuruganathan2020DGM}. Autoregressive models \cite{salimans2017PixelCNN}, variational auto-encoder (VAE) \cite{kingma2013VAE}, generative adversarial network (GAN) \cite{goodfellow2014GAN}, and denoising diffusion probabilistic model (DDPM) \cite{ho2020DDPM} are the most commonly generative models. Thereinto, GAN has been the most widely applied technique to synthesize artificial data and overcome the problem of limited data. Since the GAN-based generative model of EEG signal, i.e., EEG-GAN \cite{hartmann2018EEGGAN}, BCI researchers have successively developed several GAN-based SSVEP generation models. In 2019, Aznan et al. firstly exploited the GAN-based generative model in circumventing the limited calibration data via generating supplementary synthetic data to enlarge the size of training data \cite{aznan2019SSVEPGAN}. Only one year later, they also proposed a subject-invariant SSVEP GAN (SIS-GAN) to generate artificial EEG data that learns the subject-invariant features from the multiple SSVEP categories  \cite{aznan2021SISGAN}. Two years later, inspired by StarGAN v2, which has been used to solve multidomain image-to-image conversion, Kwon et al. proposed a novel multidomain signal-to-signal transformation method which is capable of generating artificial SSVEP signals from resting EEG \cite{kwon2022StarGAN}.

Although these GANs based on SSVEP signals have made noticeable progress, these studies only focus on generating simulated data to enlarge the amount of the training dataset, without considering the extension of signal length. However, extensive previous researches have demonstrated that longer SSVEP signal length would often achieve more accurate recognition results under the same conditions \cite{ravi2020CCNN, guney2021tsDNN, pan2022SSVEPNet, chen2023SSVEPformer}. Accordingly, in this study, we proposed a GAN-based end-to-end signal transformation network for Time-window length Extension, termed as TEGAN. TEGAN transforms short-length SSVEP signals into long-length artificial SSVEP signals. By incorporating a novel U-Net generator architecture and auxiliary classifier into the network design, the TEGAN could produce conditioned features in the synthetic data. Additionally, to regularize the training process of GAN, we introduced a two-stage training strategy and the LeCam-divergence regularization term during the network implementation. Two-stage training strategy leverages the existing source subject database to learn the common knowledge of SSVEPs for source GAN, then fine-tune to target GAN using limited target subject’s data for studying the subject-specific characteristics. LeCam-divergence regularization term regulates the discriminator predictions during the training phase via providing meaningful constraints. The proposed methods were evaluated on two public SSVEP datasets. With the assistance of TEGAN, the performance of traditional frequency recognition methods and DL-based methods have been significantly improved under limited calibration data. The extensive experimental analysis demonstrates the effectinveness of the proposed methods, while the novelty of our augmentation strategies shed some value light on understanding the subject-invariant properties of SSVEPs.

\section{Materials and methods}\label{sec2}
\subsection{Datasets}
In this study, two public SSVEP datasets were employed to evaluate the proposed augmentation methods. According to the design purpose of these two datasets, we hereinafter termed them as Direction SSVEP dataset and Dial SSVEP dataset, respectively. The specific details of each dataset are described as follows:

\subsubsection{Direction SSVEP dataset}
This dataset was published by Lee et al. in 2019 \cite{lee2019GIGADataset}. Fifty-four healthy subjects (25 females, aged 24-35 years) participated in the experiment. The experiment collected EEG data of subjects in two different periods (Session1 and Session2). The data in each period was divided into offline analysis stage and online testing stage. For the sake of simplicity, the offline data from Session1 was chosen for experimental evaluation in this study.

In the process of data acquisition, the four target stimuli were coded by 5.45 Hz, 6.67 Hz, 8.57 Hz and 12 Hz , and displayed in the lower, right, left and upper directions of the monitor, respectively. Participants were asked to focus on the center of the black screen, and then on the direction of the target stimulus highlighted in different colors. In each trial, each SSVEP stimulus was presented for 4 s, and the interval between two stimuli was 6 s. Each target frequency was presented 25 times, leading to a total of 100 trials (4 classes $\times$ 25 trial). EEG data were collected with 62 Ag/AgCl electrodes at a sampling rate of 1000 Hz were recorded in the experiment. Ten electrodes (P7, P3, Pz, P4, P8, PO9, PO10, O1, Oz and O2) covering the occipital lobe area were selected for our research. All data was down sampled to 100 Hz and band-pass filtered between 4 and 40 Hz through a fourth-order Butterworth band-pass filter. 


\subsubsection{Dial SSVEP dataset}
This open access dataset was provided by Nakanishi et al. in 2015 \cite{nakanishi2015eCCA}. In this dataset, ten healthy subjects (1 female, mean age:28 years) participated in the experiment. The subjects were instructed to sit in a comfortable chair 60 cm in front of a liquid crystal display (LCD) monitor in a dim room. The twelve flicking stimuli ($f_0=9.25$Hz, $\Delta f=0.5$Hz) were arranged in a 4$\times$3 grid space as simulation a dial interface on the monitor. The EEG data were acquired using the BioSemi ActiveTwo EEG system with a sampling rate of 2048 Hz. Eight Ag/AgCl electrodes, i.e., PO7, PO3, POZ, PO4, PO8, O1, Oz, and O2, were used during the experiment. For each subject, there were 15-run experiments. In each run, 12 trials corresponding to all 12 stimuli were generated in a random order. Thus, a total of 180 trials were collected in the experiment. Each trial was composed of 1 s cuing period and 4 s targeted flickering period. 

All data was down sampled to 256 Hz and band-pass filtered between 6 and 80 Hz through a fourth-order Butterworth band-pass filter. To suppress the adverse effect of visual latency, the data segment was extracted from the 135 ms after the stimulus onset.


\subsection{Frequency recognition methods}
In this subsection, we briefly introduce two traditional frequency recognition methods and two DL-based methods. All of these state-of-the-art methods are adopted as baselines to verify the effectiveness of the proposed methods.

\subsubsection{Traditional methods}
\begin{itemize}
	\item \textbf{IT-CCA}:In the original CCA algorithm, the sinusoidal-cosine based artificial reference signals lack the specific characteristics of subjects.  Then, IT-CCA was proposed to substitute the artificial reference signals by individual template reference signals obtained by averaging multiple training trials of specific subject \cite{bin2011ITCCA}. This strategy has been proved to be able to suppress spontaneous EEG interference and achieve better classification performance than original CCA algorithm.
	
	\item \textbf{TRCA}:TRCA is a method which learns the spatial filters to extract task related components and suppress the noise components by maximizing the reproducibility of neuroimaging data during the task period \cite{tanaka2013TRCA}. For SSVEP data, TRCA seeks to find a linear weight vector, i.e. spatial filter to maximize the inter-trial correlation of its linear combinations, referring as task-related components. In 2018, Nakanish et al. first applied TRCA algorithm to build a high-speed SSVEP-based BCI system \cite{nakanishi2017TRCA}. Nowadays, TRCA has become one of the most popular benchmark frequency recognition methods for extensive studies \cite{ravi2020CCNN, guney2021tsDNN, pan2022SSVEPNet, chen2022SSVEPformer, luo2022SAME}.
	
\end{itemize}

\subsubsection{DL-based methods}
\begin{itemize}
	\item \textbf{EEGNet}: EEGNet is a robust DL model which could yield comparable performance across multiple EEG tasks and datasets \cite{lawhern2018EEGNet, zhang20213DCNN}. It is mainly comprised of a temporal filtering layer, a depthwise convolution layer, a separable convolution layer, and a fully connected layer. Among these network components, depthwise and separable convolution make the model become compact and efficient. Due to its effectiveness, Waytowich et al. used EEGNet to decode SSVEP signals for inter-subject classification on Dial SSVEP dataset, and has achieved about 80\% accuracy \cite{waytowich2018EEGNet}.
	
	\item \textbf{C-CNN}: The frequency domain of SSVEP data contains abundant frequency and phase information relevant to the recognition task. If this information could be adequately exploited, the classification performance could be further improved. Therefore, Ravi et al. utilized FFT to transform SSVEP signals from the time domain to the frequency domain and designed a shallow CNN consisting of a spatial filter layer, a convolutional layer, and a fully connected layer to handle complex spectral data \cite{ravi2020CCNN}. C-CNN was evaluated on a 7-class dataset and the Dial SSVEP dataset, yielding satisfactory results for both intra- and inter-subject classification.
	
\end{itemize}

\subsection{The proposed augmentation methods}

\begin{figure}[htbp]
	\center{\includegraphics[scale=0.23]{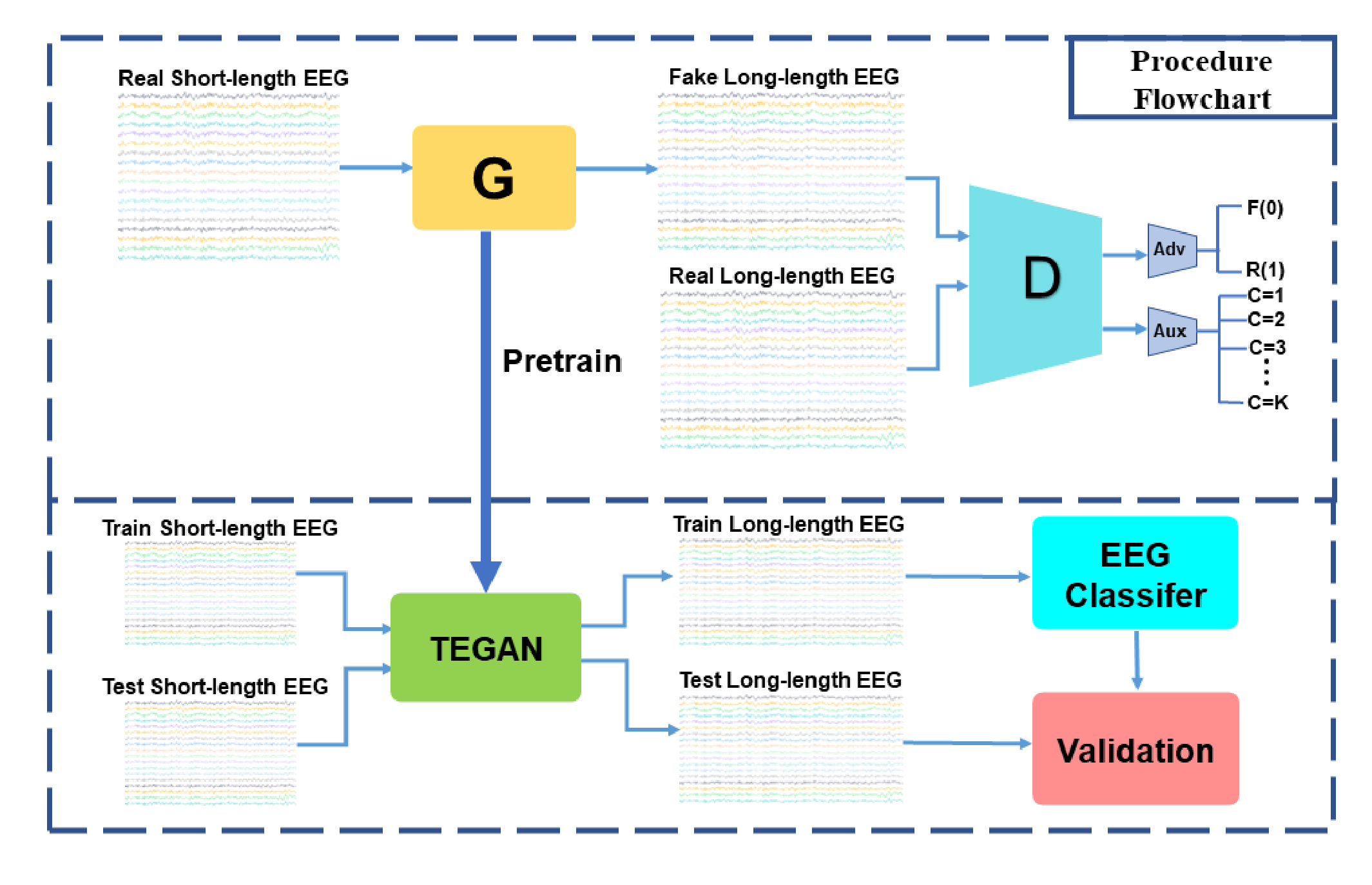}
		\caption{The flowchart of the proposed augmentation method. The whole process is divided into two steps. In the first step, real short-length EEG and real long-length EEG of training dataset are used to train the GAN model. In the second step, the pretrained generator is employed to transform all input short-length EEG into synthetic long-length EEG. Then the synthetic long-length EEG are used to train the EEG classifier and conduct classification.}
		\label{fig1}}
\end{figure}

\subsubsection{Overall framework}
The procedure flowchart of the proposed augmentation method is illustrated in Fig. \ref{fig1}. The whole process is divided into two steps. In the first step, we train the TEGAN, which could transform natural short-length EEG into artificial long-length EEG. Concretely, followed by the auxiliary classifier GAN (ACGAN) paradigm \cite{odena2017ACGAN}, TEGAN mainly includes two components, i.e., a generator and a discriminator, competing in a zero-sum game fashion. The generator receives the real short-length EEG as network input, then output the extended EEG data, namely artificial long-length EEG data. The discriminator is responsible for distinguishing between real long-length EEG data and generated long-length EEG data and simultaneously identifying their respective classes. Let $L_S$ be the likelihood that the TEGAN accurately classifies an input as real long-length EEG or synthetic long-length EEG, and let $L_C$ be the log-likelihood of the correct class:

\begin{equation}
	L_S=\mathop{\mathbb{E}}\limits_{x\sim p_{x_l}}[1-D(x)]+\mathop{\mathbb{E}}\limits_{x\sim p_{x_s}}[1+D(G(x))] 
\end{equation}

\begin{equation}
	L_C=\mathop{\mathbb{E}}\limits_{x\sim p_{x_l}}[log(D(x)\in C))]+\mathop{\mathbb{E}}\limits_{x\sim p_{x_s}}[log(D(G(x)\in C))]
\end{equation}

where $p_{x_s}$ and $p_{x_l}$ is the data distribution of short-length EEG data and long-length EEG data from the training dataset, respectively. The notation $D(x\in C)$ represents the probability of the class label being correctly identified. Let $V_G$ and $V_D$ denote the training objectives of the generator $G$ and discriminator $D$, respectively. Then the training of the TEGAN can be generally expressed as:

\begin{equation}
	\mathop{\mathrm{min}}\limits_{G}V_G=L_S-L_C
\end{equation}

\begin{equation}
	\mathop{\mathrm{min}}\limits_{G}V_G=L_S+L_C
\end{equation}

To optimize these two objective functions, we use stochastic gradient descent algorithm to train discriminator and generator alternately, and obtain the optimal parameters of generator $\theta_{G}^{*}$. Hence, a short-to-long signal converter $\zeta(\cdot)$ could be constructed using $\theta_{G}^{*}$ as follows:

\begin{equation}
	\zeta(x_s)=G(x_s|\theta=\theta_G^{*})=x_l
\end{equation}

In the second step, the converter $\zeta(\cdot)$ is employed to transform all short-length EEG into synthetic long-length EEG. Finally, the synthetic long-length EEG are used to train the EEG classifier and conduct classification. It is worth noting that since the signal converter is trained using only the training dataset and does not require label as input of pretrained generator.

\subsubsection{The architecture of generator}
\begin{figure*}[htbp]
	\center{\includegraphics[scale=0.32]{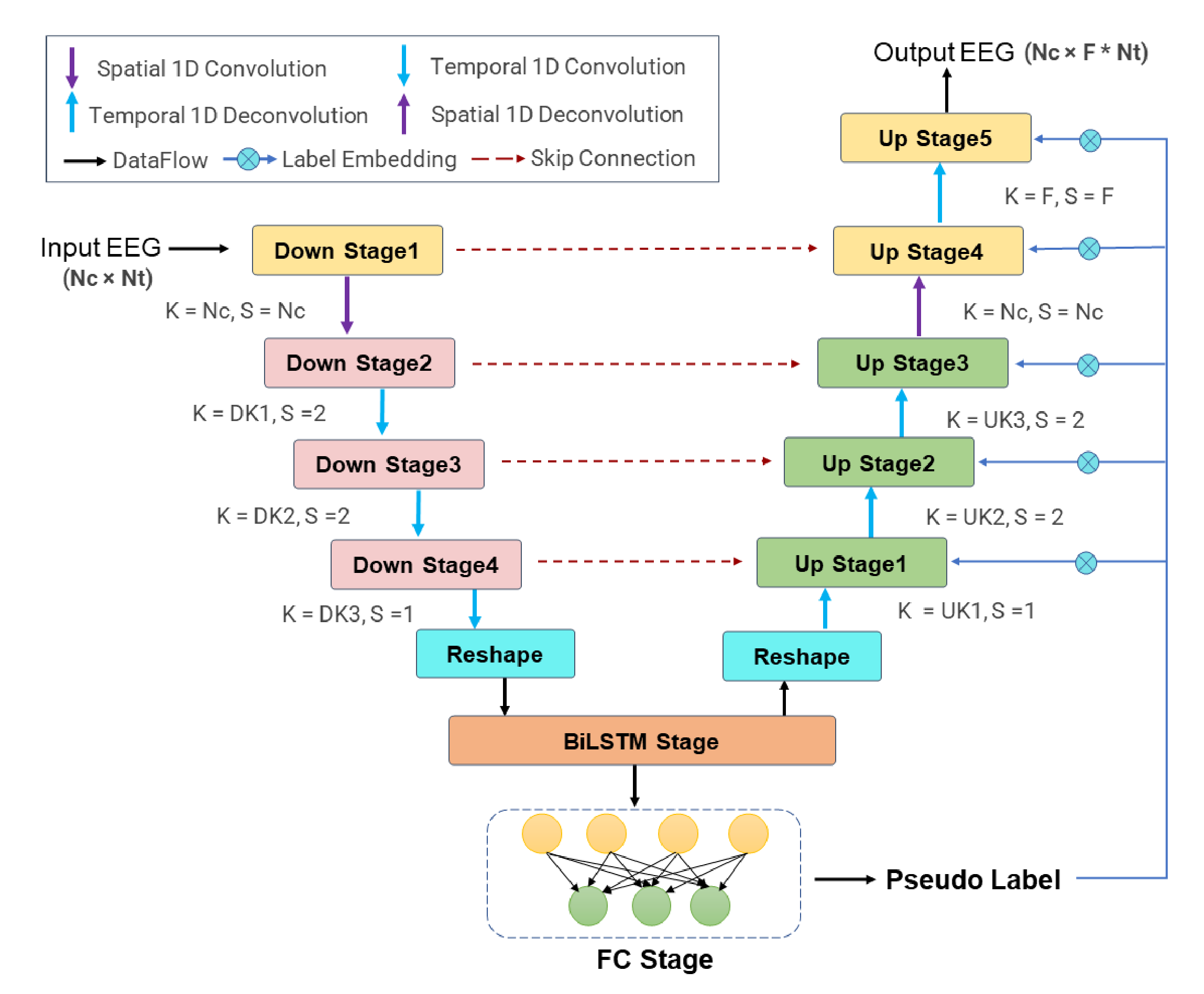}
		\caption{The architecture of generator. The whole architecture of the generator follows the design of U-Net, which is divided into two stages: down sampling and up sampling. Where Nc is the number of EEG channels, Nt denotes the number of sample points, and F indicates the extension factor. K, S represents the size of convolution/deconvolution kernel and stride, respectively. DKi, UKi (i=1, 2, 3) corresponding to kernel size for down sampling stage and up sampling stage, respectively. }
		\label{fig2}}
\end{figure*}

\begin{table*}[]
	\caption{\centering{The detailed network parameters of generator}}
	\renewcommand{\arraystretch}{1.3}
	\scalebox{0.85}{
		\begin{tabular}{c|c|c|c|c}
			\hline
			Block                                    & Module                                & Layer      & Output                 & Description                                                                                                                                                                                                               \\ \hline
			\multirow{6}{*}{Down Sampling}           & Input Module                          & Input      & $(b, 1, N_c, N_t)$     &                                                                                                                                                                                                                     
			$(batch\_size, 1, num\_channels, num\_time\_steps)$
			\\ \cline{2-5} 
			& \multirow{4}{*}{DownStage Module($\times$ 4)} & Conv1D     & $(b, DC_i, 1, DT_i)$       & \begin{tabular}[c]{@{}c@{}}kernel = $\{N_c, DK_1, DK_2, DK_3\}$ \\ stride = $\{N_c, 2, 2, 1\}$\\ $C_i = 2^{i+1} * N_c, i = \{0, 1, 2, 3\}$ \\ $DT_0$ = $N_t$, $DT_{i+1} = (DT_i - kernel{[}i{]})\ //\  stride{[}i{]} + 1$\end{tabular} \\
			&                                       & Norm       & $(b, DC_i, 1, DT_i)$       & \begin{tabular}[c]{@{}c@{}}max\_norm (= 1.0) + BN, $i = 0$ \\ SN\cite{miyato2018SN} + BN, $i = 1, 2, 3$\end{tabular}                                                                                                                             \\
			&                                       & Activation & $(b, DC_i, 1, DT_i)$       & PReLU                                                                                                                                                                                                               
			\cite{KaiMing2015PReLU}
			
			\\
			&                                       & Dropout    & $(b, DC_i, 1, DT_i)$       & $p$ = 0.5                                                                                                                                                                                                                   \\ \cline{2-5} 
			& Reshape Module                        & Reshape    & $(b, DT_3, DC_3)$          &                                                                                                                                                                                                                           \\ \hline 
			\multirow{3}{*}{BiLSTM Encoding}         & \multirow{3}{*}{BiLSTM Module}        & LSTM       & $(b, 2 * DT_3, DC_3)$      & \begin{tabular}[c]{@{}c@{}}input\_size, hidden\_size = $DC_3$, num\_layers = 1\\ batch\_first = True, bidirectional = True\end{tabular}                                                                                      \\
			&                                       & Reshape    & $(b, 1, 2 * DT_3 * DC_3)$  &                                                                                                                                                                                                                           \\
			&                                       & AvgPool    & $(b, 1, DT_3 * DC_3)$      & kernel = 2, stride = 2                                                                                                                                                                                                    \\ \hline
			\multirow{4}{*}{Pesudo Label Generation} & \multirow{3}{*}{FC Module}            & Flatten    & $(b, DT_3 * DC_3)$         &                                                                                                                                                                                                                           \\
			&                                       & Linear     & $(b, K)$                &                                                                                                                                                                                                                           \\
			&                                       & Norm       & $(b, K)$                & SN                                                                                                                                                                                                                        \\ \cline{2-5} 
			& Selection Module                      & Argmax     & $(b, 1)$                 & select the highest probability as pseudo label                                                                                                                                                                            \\ \hline
			\multirow{8}{*}{Up Sampling}             & Reshape Module                        & Reshape    & $(b, DC_3, 1, DT_3)$       &                                                                                                                                                                                                                           \\ \cline{2-5} 
			& \multirow{6}{*}{UpStage Module($\times$ 5)}   & DeConv1D   & $(b, UC_i, 1, UT_i)$       & \begin{tabular}[c]{@{}c@{}}kernel = \{$UK_1, UK_2, UK_3, N_c, F$\}\\ stride = \{$1, 2, 2, N_c, F$\}\\ UC = \{$8 * N_c, 8 * N_c, 4 * Nc, 2 * N_c, N_c$\}\\ UT = \{$DT_2, DT_1, DT_0, N_t, 2 * N_t$\}\end{tabular}                                \\
			&                                       & Norm       & $(b, UC_i, 1, UT_i)$       & SN + cBN                                                                                                                                                                                                                 
			\cite{dumoulin2016CBN}
			\\
			&                                       & Activation & $(b, UC_i, 1, UT_i)$       & PReLU                                                                                                                                                                                                                     \\ \cline{3-5} 
			&                                       & Conv1D     & $(b, UC_i \ // \  M_i, 1, UT_i)$ & \begin{tabular}[c]{@{}c@{}}kernel = 1, stride = 1, $M = \{0, 2, 2, 2, N_c\}$\\ only exist when $i \geq 1$\end{tabular}                                                                                              \\
			&                                       & Norm       & $(b, UC_i \ // \  M_i, 1, UT_i)$ & SN + cBN, only exist when $i \geq 1$                                                                                                                                                                             \\
			&                                       & Activation & $(b, UC_i \ // \  M_i, 1, UT_i)$ & PReLU, only exist when $i \geq 1$                                                                                                                                                                               \\ \cline{2-5} 
			& Output Module                         & Output     & $(b, 1, N_c, F * N_t)$     &                                                                                                                                                                                                                           \\ \hline
		\end{tabular}
	}
	\label{tab1}
\end{table*}

The detailed architecture of generator is presented in Fig.~\ref{fig2}. The whole architecture of the generator follows the U-Net architecture proposed by Ronneberger et al. \cite{ronneberger2015UNet}, which is divided into down sampling stages and up sampling stages. Moreover, considering the dependency between spatial-temporal features of EEG data, we utilize a bidirectional long short-term memory (Bi-LSTM) network to encode the features derived by the last down-sampling module, thus connecting the down sampling stages and up sampling stages. Furthermore, we added a conditional batch normalization (cBN) layer in each up-sampling module to mitigate the adverse effects of homogenization between different synthetic samples \cite{dumoulin2016CBN}. However, due to the inaccessibility of label information in the input end of generator, we adopted a fully connected layer to create high-confidence pseudo label, which can be improved via minimizing the following objective function $L_G$:

\begin{equation}
	\mathop{\mathrm{min}}\limits_{G}L_G=V_G - \sum_{c=1}^{K}y_c\log{p_c}
\end{equation}

where $y_c$ is realistic label with one-hot encoding, $p_c$ is the predicted probability of class $c$, and $K$ is number of classes. 

The network parameters of generator are listed in Table~\ref{tab1}. Among the parameters, the kernel size $DK$ for down-sampling module could be determined through the following formula:

\begin{equation}
	DK_i = \lceil D_i * ws \rceil, i = 1, 2, 3
\end{equation}

where $D = \{20, 12, 8\}$, and $ws$ represents the window size of input short-length EEG signal. Then the kernel size $UK$ for up-sampling module could be calculated as:	

\begin{equation}
	UK_{i+1} = DT_{2-i} - (DT_{3-i}) * S_i , i = 0, 1, 2
\end{equation}

Here, $S = \{1, 2, 2\}$, and DT represents the number of time steps for down sampling module, which is presented in Table~\ref{tab1}.  

\subsubsection{The architecture of discriminator}

\begin{figure*}[htbp]
	\center{\includegraphics[scale=0.28]{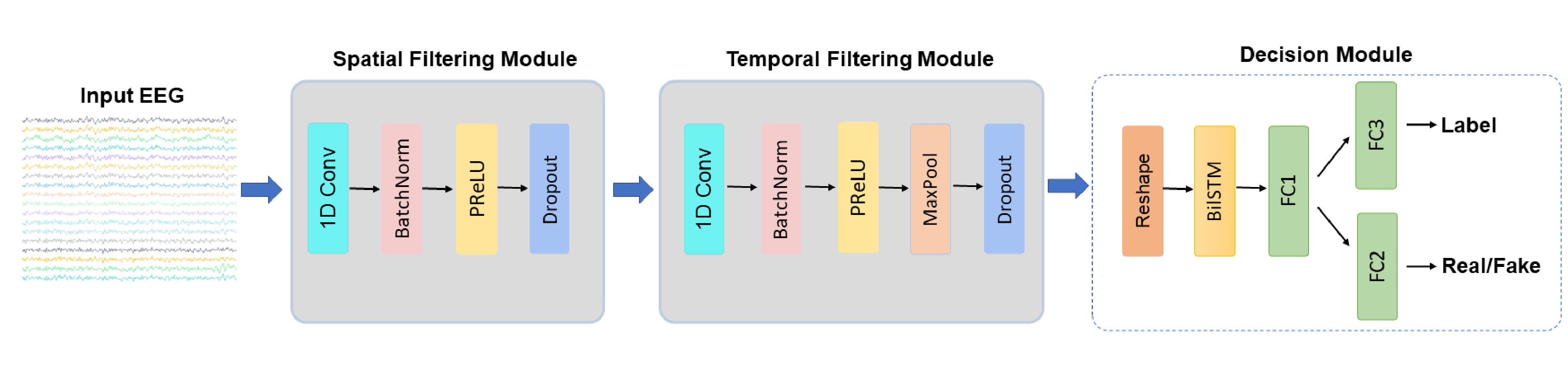}
		\caption{The architecture of discriminator. The discriminator follows the design of SSVEPNet, which mainly comprised of a spatial filtering module, a temporal filtering module, and a decision module. }
		\label{fig3}}
\end{figure*}

The minute architecture of discriminator is shown in Fig. ~\ref{fig3}. Overall, the discriminator follows the design of our previously proposed SSVEPNet model \cite{pan2022SSVEPNet}. It is mainly comprised of three modules, namely spatial filtering module, temporal filtering module, and decision module. In the spatial filtering module, a one-dimensional convolution (1D Conv) layer is used to fuse different channel information of EEG. In the temporal filtering module, 1D Conv layer is employed to extract temporal features of EEG. In the decision module, a Bi-LSTM layer is leveraged to learn the dependence between spatial-temporal features, while fully connected layers are employed for classification and authentication. The distinction between SSVEPNet and the proposed discriminator only lies in two points. One is that a max pooling layer with a kernel size of 2 has been added to the temporal filtering module. Another is that the fully connection layer in the decision module has changed from three layers to two layers, and the neurons in the first layer is equal to twentieth of encoded spatial-temporal features outputted by BiLSTM layer. The remaining network parameters of the discriminator are identical with SSVEPNet.

\subsubsection{Regularizing TEGAN on limited data}
GAN is notoriously difficult to train, it is highly susceptible to encounter with mode collapse phenomenon - particularly from datasets with high variability. Furthermore, this problem may deteriorate with a limited training data. Generators are inclined to merely memorize limited training samples, rather than learning the sophisticated data distribution of training dataset  \cite{odena2017ACGAN}. To moderate this issue, we introduce a two-stage training strategy and LeCam divergence regularization term to regularize the training process of TEGAN.

Firstly, inspired by the previous study \cite{guney2021tsDNN}, we utilized transfer learning technique to design a two-stage training strategy for the proposed GAN framework. Suppose there are $N$ subjects in total, in the first stage, we trained two global models $D_s$ and $G_s$ using the $N-1$ source subject data i.e. cross-subject. In the second stage, the network parameters of $D_s$ and $G_s$ were directly duplicated to two target models, $D_t$ and $G_t$. Then we froze the parameters in $D_t$  in addition to the fully connection layers, and fine-tune $D_t$ and $G_t$ using a limited amount of training data of target subjects. 

Secondly, we supplemented a LeCam regularization term in the training objective of discriminator, which is used to regulate the discriminator predictions during the training phase \cite{tseng2021LeCam}:

\begin{equation}
	\mathop{\mathrm{min}}\limits_{D}L_D = -V_D + \lambda R_{LC}(D)
\end{equation}

where $\lambda$ represents regularization weight, and LeCam regularization term $R_{LC}(D)$ is expressed as:

\begin{equation}
	R_{LC}(D)=\mathop{\mathbb{E}}\limits_{x\sim p_{x_l}}[{\Vert D(x)-\alpha_F \Vert}^2]-\mathop{\mathbb{E}}\limits_{x\sim p_{x_s}}[{\Vert D(G(x))-\alpha_R \Vert}^2]
\end{equation}

where $\alpha_F$ and $\alpha_R$ are anchors obtained by the exponential moving average variables, which is aiming at tracking the discriminator predictions. $R_{LC}(D)$ pushes the discriminator to mix the predictions of real and generated EEG, rather than differentiating them. The counterintuitive effect could serve as regularization by offering meaningful constraints for optimizing a more robust objective. Assuming that there are $TE$ training epochs in total, then $\alpha_F$ and $\alpha_R$ could be calculated using the following equations:

\begin{equation}
	\left\{\begin{aligned}
		\alpha_F(i+1)=\gamma D(G(x_s))+(1 - \gamma)\alpha_F(i) \\
		\alpha_R(i+1)=\gamma D(x_l)+(1 - \gamma)\alpha_R(i)
	\end{aligned}\right.
	\  ,SE \leq i \leq TE
\end{equation}

Here, $\gamma$ is decay coefficient, and SE represents the starting epoch to implement LeCam regularization term, which is conducive to avoid the excessive regularization in the initial stage with under fitting. 

\subsection{Experimental evaluation}

To validate the efficacy of the proposed method, we conducted intra-subject classification and ablation experiments, and adopted above four traiditional and DL-based methods as the evaluated classifiers.  

The EEG data from $N-1$ source subject was divided into short-length training set $\mathcal{T}_{tr} (X_s^s)$ and its corresponding long-length training set $\mathcal{T}_{tr} (X_l^s)$. The target subject data was split into short-length training set $\mathcal{T}_{tr} (X_s^t)$, long-length training set $\mathcal{T}_{tr} (X_l^t)$ and short-length testing set $\mathcal{T}_{tt} (X_s^t)$. Among them, $\mathcal{T}_{tr} (X_s^t)$ and $\mathcal{T}_{tr} (X_l^t)$ have equal quantities and correspond one-to-one. As illustrated in Table~\ref{tab2}, $\mathcal{T}_{tr} (X_s^t)$ and $\mathcal{T}_{tt} (X_s^t)$ were divided into three proportions: 2:8, 5:5 and 8:2.  We marked these three proportions of datasets as large-scale, medium-scale and small-scale dataset. We exploited the two-stage training stage using $\mathcal{T}_{tr} (X_s^s)$, $\mathcal{T}_{tr} (X_l^s)$, $\mathcal{T}_{tr} (X_s^t)$, $\mathcal{T}_{tr} (X_l^t)$ to optimize the parameters of source generator $G_s$ and target generator $G_t$, and then employed $G_t$ to transform the $T_{tr} (X_s^t )$,  $T_{tt} (X_s^t )$ into $\mathcal{T}_{tr} (X_{\hat{l}}^t)$, $\mathcal{T}_{tt} (X_{\hat{l}}^t)$, respectively. Finally, the $\mathcal{T}_{tr} (X_{\hat{l}}^t)$ was used to train SSVEP classifiers, and undertook evaluation on $\mathcal{T}_{tt} (X_{\hat{l}}^t)$. To eliminate the randomness of data partitioning, the K-Fold evaluation strategy was adopted in the experiments, under which the parameters of $G_s$ remained unchanged while $G_t$ was updated K times in the whole process.  

With respect to ablation experiments, we compared the classification performance with and without four pivotal components, i.e. auxiliary classifier in the discriminator, the pseudo-label generation in the generator, the two-stage training strategy, and the LeCam divergence regularization term. The traiditional methods and DL-based methods were verified on small sample size calibration data in the intra-subject experimental paradigm. It is noteworthy that after removing the two-stage training strategy, the proposed TEGAN model would only be trained by target subject data. The expansion factor $F$ is set to 2 by default in both intra-subject and ablation experiments. However, for the purpose of maximizing the effectiveness of TEGAN as much as possible, we also explore the impact of $F=3, 4$ on the performance of TEGAN in the subsequent analysis.

In this study, we implemented the proposed augmentation method in PyTorch framework. The trained and tested models were running on a server with an Intel Core I7-10700K CPU and an NVIDIA GeForce GTX 3090 (24GB Memory) GPU. The implementation code would be available at \url{https://github.com/YudongPan/TEGAN}. The hyperparameters on two datasets were set as follows.

\begin{itemize}
	\item \textbf{Direction SSVEP dataset}: For the implementation of the proposed GAN model, in the first training stage, mini-batch (B) = 64, epochs (E) = 200, learning rate (lr) = 0.001, optimizer (Opt) = Adam (beta1=0.9, beta2=0.999), weight decay (wd) = 0.0001. In the second training stage, B = 20, E = 500, lr = 0.01, Opt = Adam (beta1=0.9, beta2=0.999) + Cosine Annealing, wd = 0.0003. For the LeCam divergence regularization term, SE = 50, $\lambda$ = 0.6, $\gamma$ = 0.90. For the implementation of DL-based recognition methods in the intra-subject experiments, B = 20, E = 200, lr = 0.001, wd = 0.0001.   
	
	\item \textbf{Dial SSVEP dataset}:For the first and second training stage, all hyperparameters are identical with Direction SSVEP dataset, except for the B = 24 in the second training stage. For classification experiments of DL-based methods, all hyperparameters remained unchanged as Direction SSVEP Dataset, except B = 30 on intra-subject experiments. 
\end{itemize}

\subsection{Statistical analysis}
In this study, classification accuracy and ITR were adopted as the evaluation metrics. The classification accuracy $ P\in [0,1]$ is the proportion of the correct classification of all EEG samples tested, and the ITR is calculated by the following formula. 
\begin{equation}
	ITR(P,T)=(log_2N+Plog_2P+log_2\frac{(1-P)}{(N-1)}) \times \frac{60}{T}    
\end{equation}

Where $T$ is time window length used for evaluation, and $N$ denotes number of stimulus targets. During calculating the ITR, we take a 0.5 s gaze shift time into account, complying with the previous studies \cite{zhang2018CORCA, chen2022SSVEPformer}. The averaged classification accuracy and ITR across all subjects for K-folds validation were presented in the form of mean $\pm$ standard deviation. This study calculated these two measurements on the averaged level with different data lengths (0.5 s to 1 s, with a step of 0.1 s). Paired-sample t-test was implemented to investigate whether there was significant difference in the classification accuracy or ITR between each pair of methods at each condition.

\begin{table*}[]
	\caption{Data proportions of short-length target subject data}
\begin{center}
		\begin{tabular}{ccccccc}
			\hline
			\rule{0pt}{10pt}
			\multirow{2}{*}{Data Amount} & \multicolumn{1}{l}{} & \multicolumn{2}{c}{Direction SSVEP Dataset}                                             & \multicolumn{1}{l}{} & \multicolumn{2}{c}{Dial SSVEP Dataset}                                             \\ \cline{3-4} \cline{6-7}
			\rule{0pt}{10pt}
			& \multicolumn{1}{l}{} & \multicolumn{1}{l}{Split Ratio} & \multicolumn{1}{l}{Trial Distribution} & \multicolumn{1}{l}{} & \multicolumn{1}{l}{Split Ratio} & \multicolumn{1}{l}{Trial Distribution} \\ \hline
			\rule{0pt}{8pt}
			Large                        &                      & 8:2                             & 80:20                                &                      & 8:2                             & 144:36                                  \\
			\rule{0pt}{8pt}
			Middle                       &                      & 5:5                             & 50:50                                  &                      & 5:5                             & 84:84                                  \\
			\rule{0pt}{8pt}
			Small                        &                      & 2:8                             & 20:80                                &                      & 2:8                             & 36:144                                  \\ \hline
		\end{tabular}
\end {center}
\label{tab2}
\end{table*}

\section{Results}
\subsection{Results of intra-subject classification experiments}
\subsubsection{Study on the impact of time-window length on the performance of the proposed method}

\begin{figure*}[htbp]
	\center{\includegraphics[scale=0.18]{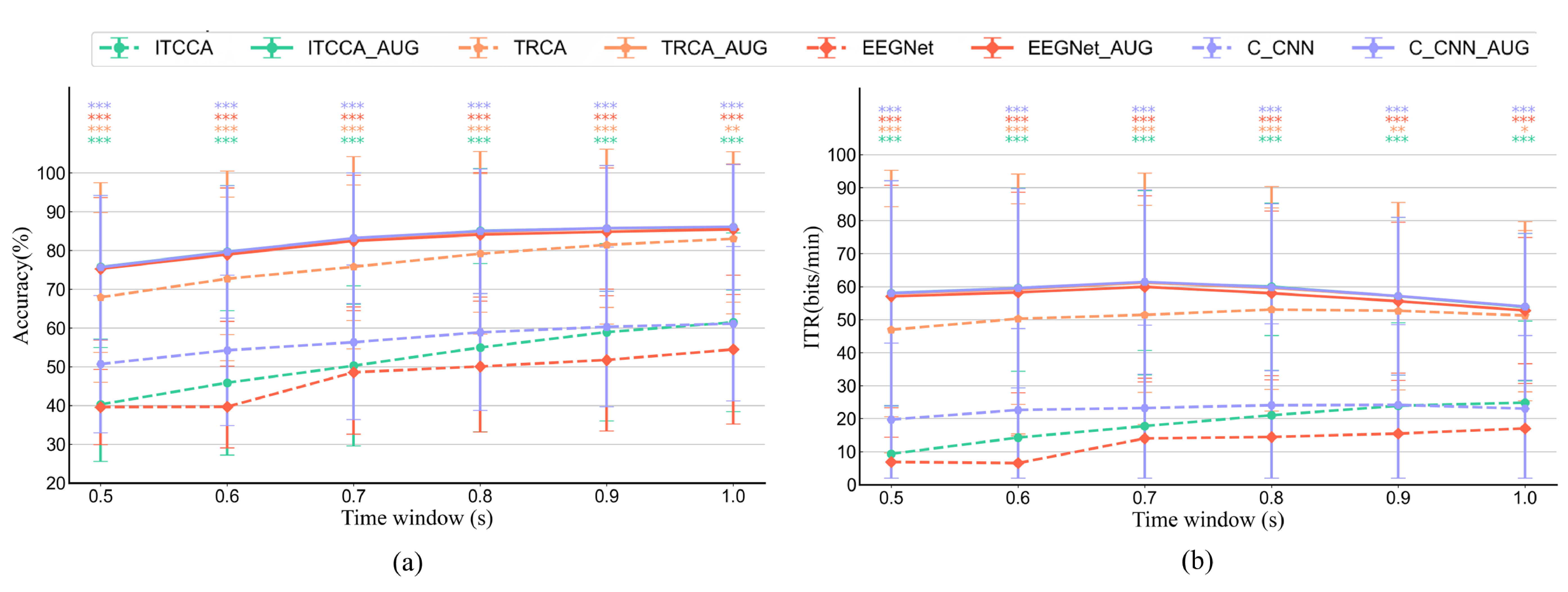}
		\caption{The classification results of original signals and augmented signals across all subjects on small-scale dataset of Direction SSVEP Dataset obtained by four baseline classification methods as ITCCA, TRCA, EEGNet and C-CNN at various time window lengths. (a) average accuracies, (b) average ITRs. The time window length of the real short-length EEG increases from 0.5 s to 1.0 s, corresponding to the generated long-length EEG increased from 1.0 s to 2.0 s. AUG is the result of using generated long-length EEG. The colored asterisk indicates significant difference between original signals and augmented signals evaluated  by paired t-tests (*$p < 0.05$, **$p < 0.01$, ***$p < 0.001$). }
		\label{fig4}}
\end{figure*}

\begin{figure*}[htbp]
	\center{\includegraphics[scale=0.18]{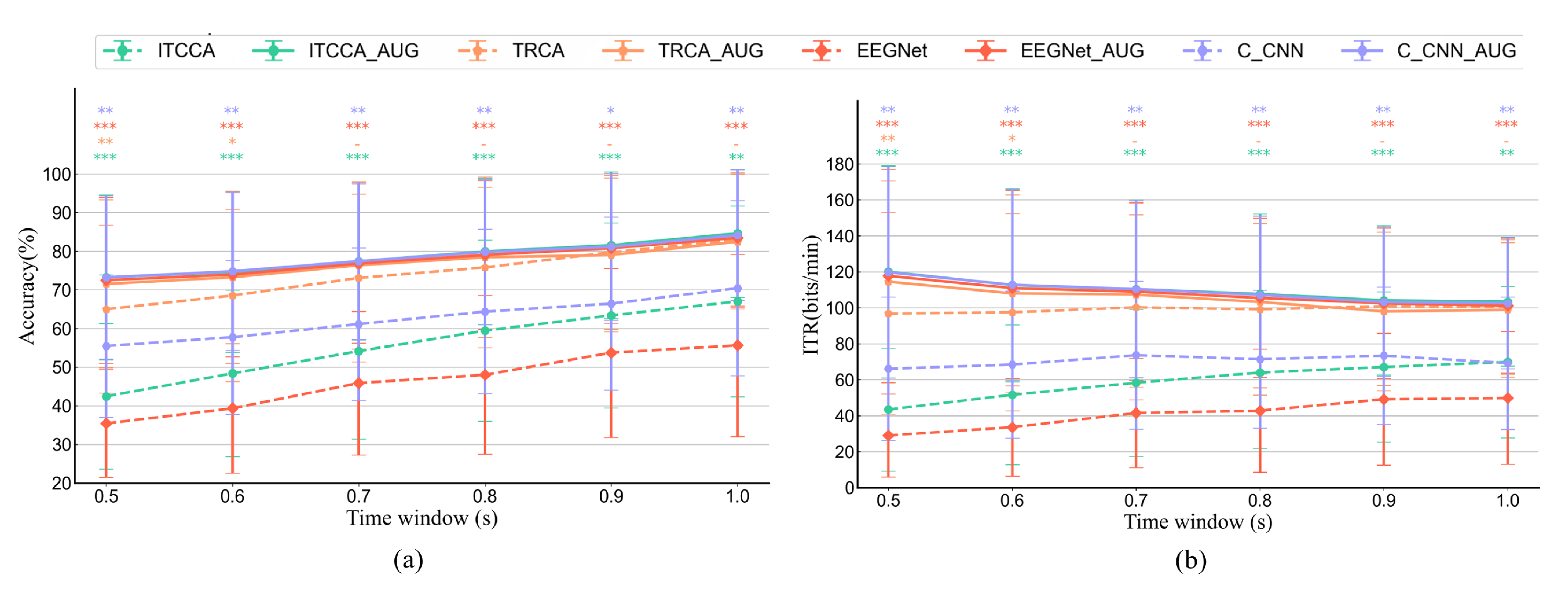}
		\caption{The classification results of original signals and augmented signals across all subjects on small-scale dataset of Direction SSVEP Dataset obtained by four baseline classification methods as ITCCA, TRCA, EEGNet and C-CNN at various time window lengths. (a) average accuracies, (b) average ITRs. The time window length of the real short-length EEG increases from 0.5 s to 1.0 s, corresponding to the generated long-length EEG increased from 1.0 s to 2.0 s. AUG is the result of using generated long-length EEG. The colored asterisk indicates significant difference between original signals and augmented signals evaluated  by paired t-tests (*$p < 0.05$, **$p < 0.01$, ***$p < 0.001$). }
		\label{fig5}}
\end{figure*}

\begin{figure*}[htbp]
	\center{\includegraphics[scale=0.18]{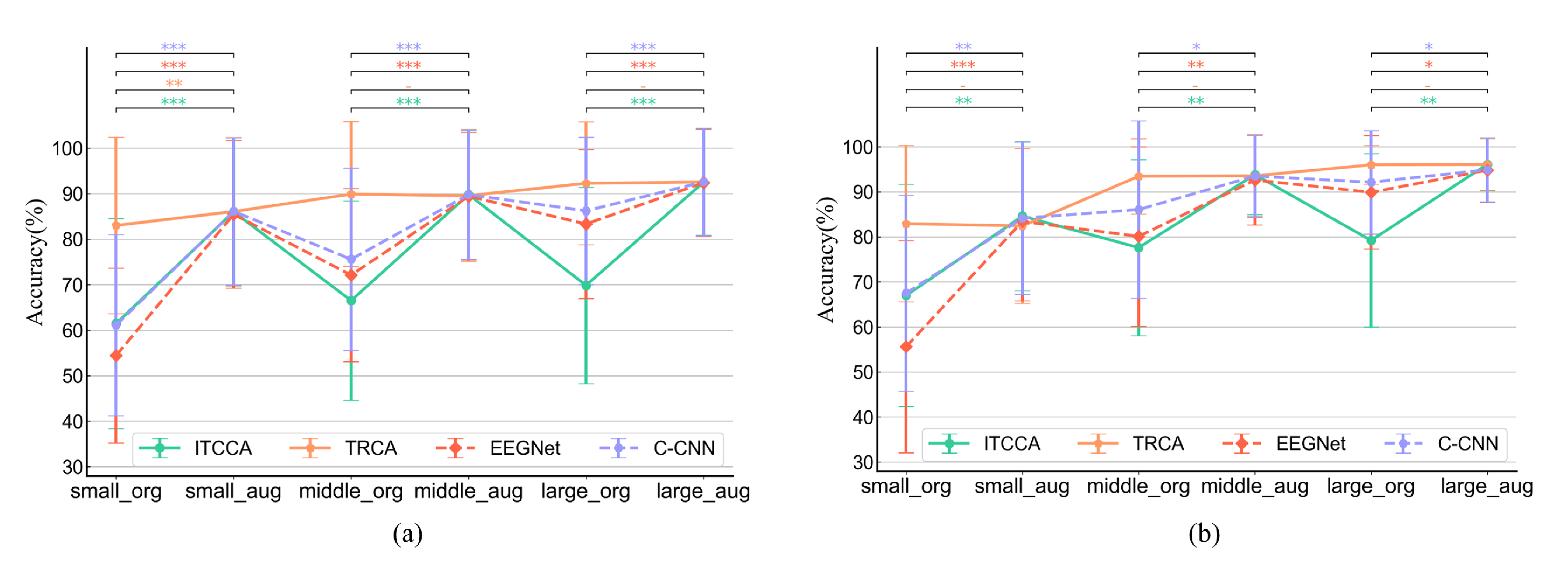}
		\caption{Averaged classification results across subjects of four baseline classification methods as ITCCA, TRCA, EEGNet and C-CNN using original signals and augmented signals at different scales of datasets on (a) Direction SSVEP Dataset, (b) Dial SSVEP Dataset. The time window length of the real short-length EEG is 1.0 s, corresponding to the generated long-length EEG at 2.0 s. ORG and AUG is the result of using real short-length EEG and generated long-length EEG, respectively. The colored asterisk in the figure indicates significant difference between original signals and augmented signals used for the same method by paired t-tests (*$p < 0.05$, **$p < 0.01$, ***$p < 0.001$). }
		\label{fig6}}
\end{figure*}

Firstly, we investigated how the classification results of four baseline methods vary with different time windows on two SSVEP datasets. In this setting, the SSVEP dataset from a target subject was divided into the training dataset and testing dataset in the portion of 2:8. The time window of the original signal ranges from 0.5 s to 1.0 s with an interval of 0.1 s, corresponding to the augmented signal increased from 1.0 s to 2.0 s with an interval of 0.2 s. Fig.~\ref{fig4} and Fig.~\ref{fig5} delineate the averaged classification accuracies and ITRs of original signals and augmented signals with six signal lengths across all subjects on Direction SSVEP Dataset and Dial SSVEP Dataset, respectively. 

For the results on Direction SSVEP dataset, it could be seen from Fig.~\ref{fig4} that the classification performance of the four baseline methods using augmented signals outperforms that using original signals at each time window. All frequency recognition algorithms could achieve higher classification accuracy and ITR after using augmented signals. For the classification accuracy, the paired t-tests showed that there were statistically significant differences between all pairs of methods across all time-window lengths (0.5 s to 0.9 s: all $p$ < 0.001, 1.0 s: all $p$ < 0.01). From 0.5 s to 1 s, the three methods ITCCA, EEGNet and C-CNN could improve the classification accuracy by more than 20\%. Especially for EEGNet, after using the augmented signal, it could improve the classification accuracy by 35.67\%, 39.28\%, 33.88\%, 34.01\%, 33.09\% and 31.00\%, respectively. As for the TRCA, its improvement effect was weaker than the other three algorithms, while there was 7.71\%, 6.69\%, 7.28\%, 5.63\%, 4.27\% and 3.05\% accuracy improvement after using the augmented signal, respectively. For the ITR, the paired t-tests revealed that there were statistically significant differences between all pairs of methods across all time windows (0.5 s to 0.8 s: all $p$ < 0.001, 0.9 s: all $p$ < 0.01, 1.0 s: all $p$ < 0.05). Similar to the classification accuracy, the improvement effect of ITCCA, EEGNet and C-CNN is more obvious than that of TRCA. Surprisingly, under all time windows, the ITR of the three methods, i.e. ITCCA, EEGNet and C-CNN, using the augmented signal, had increased by nearly 2 times compared with the ITR of the original signal. In addition, the time window corresponding to the highest ITR varied among different methods while using the original signal. Namely, the following results were achieved: ITCCA: 24.83$\pm$24.74 bits/min at 1.0 s, TRCA: 53.05$\pm$30.75 bits/min at 0.8 s, EEGNet: 17.05$\pm$19.64 bits/min at 1.0 s, C-CNN: 24.16$\pm$24.35 bits/min at 0.9 s. Whereas, the highest ITR of all methods after using the augmented signal were all achieved when the time window of 0.7 s, and the ITR results become as:ITCCA: 61.19$\pm$27.93 bits/min, TRCA: 61.18$\pm$28.03 bits/min, EEGNet: 59.92$\pm$27.66 bits/min, C-CNN: 59.84$\pm$25.30 bits/min.   

Considering that Dial SSVEP Dataset have more stimulus categories and fewer subjects, the results obtained in Dial SSVEP Dataset are different with Direction SSVEP Dataset. As illustrated in Fig.~\ref{fig5}, compared with the methods using original signal, the methods using augmented signal had achieved better classification accuracies and ITRs across all time windows, except for TRCA at 0.9 s and 1.0 s. For the classification accuracy, the paired t-tests showed that there were statistically significant differences among three pairs of methods (ITCCA vs. ITCCA\_AUG, EEGNet vs. EEGNet\_AUG, C-CNN vs. C-CNN\_AUG) across all time window (0.5 s to 0.8 s and 1.0 s: all $p$ < 0.01, 0.9 s: all $p$ < 0.05). As for the TRCA algorithm, there was a significant difference when the time-window length is 0.5 s ($p$=0.002) and 0.6 s ($p$=0.02), but there was no significant difference when the time-window length is 0.7 s to 1.0 s ($p$ > 0.05). From 0.5 s to 1.0 s, the classification accuracies improved by each algorithm were as follows: ITCCA: 30.8\%, 26.27\%, 23.25\%, 20.48\%, 18.16\%, 17.60\%; TRCA: 6.54\%, 4.72\%, 3.27\%, 2.63\%, -0.72\%, -0.49\%; EEGNet: 37.04\%, 34.61\%, 30.95\%, 31.00\%, 27.05\%, 27.80\%; C-CNN: 17.75\%, 17.05\%, 16.26\%, 15.42\%, 14.72\%, 13.73\%. For the ITR, the results are similar to the classification accuracy, showing a consistent trend. Other than TRCA, the paired t-tests revealed that there were statistically significant differences between remaining three pairs of methods across all time-window lengths (0.5 s to 1.0 s: all $p$ < 0.01). After using the augmented signal, the TRCA algorithm could significantly achieve higher ITR than that using the original signal when time-window length is 0.5 s ($p$=0.003) and 0.6 s ($p$=0.03). However, there was no significant difference when the time-window length is 0.7 s to 1.0 s. As for this dataset, the highest ITR was achieved using original signal for four baseline methods was as follows: ITCCA: 69.87$\pm$42.14 bits/min at 1.0 s, TRCA: 100.93$\pm$44.11 bits/min at 0.9 s, EEGNet: 49.82$\pm$36.97 bits/min at 1.0 s, 73.63$\pm$41.10 bits/min at 0.7 s. Whereas, the highest ITR of all methods after using the augmented signal were all achieved when the time-window length is 0.5 s, and the result was as follows: ITCCA: 120.03$\pm$58.94 bits/min, TRCA: 114.49$\pm$56.52 bits/min, EEGNet: 117.72$\pm$59.17 bits/min, C-CNN: 119.72$\pm$58.97 bits/min.

\subsubsection{Study on the impact of the scale of training dataset on the performance of proposed method}
Furthermore, we investigated the results how the classification performance of four baseline methods varied with different scales of training dataset on two SSVEP datasets. Specifically, on each dataset, the original signal length was set to 1.0 s, corresponding to its augmented signal length 2.0 s. The SSVEP dataset from target subject was divided into the training dataset and testing dataset in the portion of 2:8, 5:5 and 8:2 respectively. In this study, we marked these three scales of datasets as small-scale, middle-scale and large-scale datasets, respectively. Fig.~\ref{fig6} illustrates averaged classification results across subjects of four baseline classification methods using original signals and augmented signals at different scales of datasets on two SSVEP datasets.

From the classification results in Fig.~\ref{fig6}, we could observe that the average classification accuracy of each baseline algorithm for the original signal and the augmented signal had been gradually improved with the continuous increase of training data, and the trend is consistent on both datasets. Except for the TRCA, the paired t-tests showed that there was statistically significant difference among other three pairs of methods across three scales of training dataset on two SSVEP datasets (Direction SSVEP dataset: all $p$ < 0.001, Dial SSVEP dataset: all $p$ < 0.05). For the TRCA algorithm, the classification accuracy of the augmented signal was only significantly different from that of the original signal in the small-scale training dataset on the Direction SSVEP dataset ($p$=0.004). This indicates that augmented signal improves the classification performance on a small-scale training dataset of fewer stimulus categories more significantly. Moreover, we could find that the improvement effect of classification performance of augmented signals was gradually weakened with continuous expansion of training data. For instance, on the small-scale dataset of Direction SSVEP dataset, the improved accuracies of all recognition methods as ITCCA, TRCA, EEGNet, C-CNN were 24.54\%, 3.05\%, 31.00\%, and 24.97\%, respectively. However, the improved accuracies decreased to 23.29\%, -0.33\%, 17.31\%, 14.19\% on middle-scale dataset, and 22.63\%, 0.30\%, 9.04\%, and 6.34\% on large-scale dataset, respectively.

\subsection{Ablation experiments}
We performed ablation experiments on two SSVEP datasets to explore the contribution of pivotal components of the implemented the augmentation framework. In this experiment, the SSVEP dataset from a target subject was divided into the training dataset and testing dataset in the portion of 2:8. And the original signal length was set to 0.5 s, corresponding to its augmented signal length 1.0 s. The averaged classification accuracies across all subjects of the four baseline methods on two datasets are listed in the Table~\ref{tab3}. 

From the results in Table~\ref{tab2}, it could be observed that only two of components as Pseudo Label and Two Stage in the Direction SSVEP Dataset are advantageous to improve the classification results of each algorithm, while all components in the Dial SSVEP Dataset are all conducive to the final classification results. On the one hand, the two components of Auxiliary Classifier and LeCam Divergence do not contribute to the classification results of each frequency recognition algorithm on the 4-class dataset, i.e. Direction SSVEP Dataset, but could significantly improve the classification results of each algorithm on the 12-class dataset, i.e. Dial SSVEP Dataset (Auxiliary Classifier: all $p$<0.05, LeCam Divergence: all $p$<0.01). This implies that Auxiliary Classifier and LeCam Divergence are more conducive to inhibiting the collapse of GAN production mode and improving the richness and quality of generated data on SSVEP datasets with a large number of categories. On the other hand, the two components of Pseudo Label and Two Stage can significantly improve the classification results of each algorithm on the Direction SSVEP Dataset (Pseudo Label: all $p$<0.001, Two Stage: all $p$<0.001) and the Dial SSVEP Dataset (Pseudo Label: all $p$<0.001, Two Stage: all $p$<0.05). In addition, we found that when Pseudo Label was removed from the proposed TEGAN model, the performance difference between the frequency recognition algorithms on the two datasets gradually expanded. When we removed the Auxiliary Classifier from the TEGAN model, the classification performance gap between the two frequency recognition algorithms was only could be found on the Dial SSVEP Dataset. This indicates that Pseudo Label could provide more robust category information embedded in the generated data for the GAN model when only a small amount training data are available, as compared with the Auxiliary Classifier. In addition, since the component Two Stage could assist the GAN model improve considerable accuracy results on both datasets, we could conclude that the strategy of transfer learning could be applied to the GAN model that generates SSVEP data as well. The generalization performance of GAN could be improved leveraging the two-stage training strategy to study both subject-invariant and subject-specific characteristics.

\begin{table*}[]
	\caption{Ablation study on two Datasets. The case (a)-(e) and case (f)-(j) represent the classification results in small-scale dataset of Direction SSVEP dataset and Dial SSVEP dataset, respectively. The original signal length was set to 0.5 s, corresponding to its augmented signal length 1.0 s. The highest accuracy (in bold) indicates the optimal scheme for this method.}
	\renewcommand{\arraystretch}{1.5}
	\begin{threeparttable}
		\scalebox{0.8}{
			\begin{tabular}{cccccclllll}
				\cline{1-10}
				& \multicolumn{4}{c}{Module Selction}                                &  & \multicolumn{4}{c}{Accuracy(\%)}                                                          &  \\ \cline{2-5} \cline{7-10}
				Case & Auxiliary Classifier & Pseudo Label & Two Stage & LeCam Divergence &  & ITCCA                & TRCA                 & EEGNet               & C-CNN                &  \\ \cline{1-10}
				(a)  & \textendash       & \checkmark      & \checkmark   & \checkmark               &  & 75.46$\pm$18.66  & 75.59$\pm$18.59   & 74.93$\pm$18.77   & 75.75$\pm$18.44        &  \\
				(b)  & \checkmark   & \textendash          & \checkmark   & \checkmark                &  & 74.48$\pm$18.82*** & 74.20$\pm$19.01*** & 46.80$\pm$14.88***   & 74.43$\pm$18.59***          &  \\
				(c)  & \checkmark   & \checkmark  & \textendash      & \checkmark                &  & 59.60$\pm$22.46***    & 59.57$\pm$22.48*** & 58.64$\pm$22.20***  & 59.50$\pm$22.51***           &  \\
				(d)  & \checkmark   & \checkmark  & \checkmark  & \textendash               		 &  & \textbf{76.03$\pm$18.42}    & \textbf{76.01$\pm$18.42}     & \textbf{75.69$\pm$18.39}  & \textbf{76.02$\pm$18.43}          &  \\
				(e)  & \checkmark   & \checkmark  & \checkmark  & \checkmark                &  & 75.70$\pm$18.54  & 75.63$\pm$18.59 & 75.27$\pm$18.39  & 75.65$\pm$18.59 &  \\ \cline{1-10}
				(f)  & \textendash       & \checkmark      & \checkmark   & \checkmark               &  & 69.50$\pm$21.85**  & 56.40$\pm$16.19***   & 69.29$\pm$22.04**   & 69.09$\pm$22.76*         &  \\
				(g)  & \checkmark   & \textendash          & \checkmark   & \checkmark                &  & 65.42$\pm$21.56*** & 60.82$\pm$19.18*** & 57.23$\pm$23.35***   & 68.22$\pm$23.14***           &  \\
				(h)  & \checkmark   & \checkmark  & \textendash      & \checkmark                &  & 61.86$\pm$25.56*    & 64.99$\pm$24.28*  & 66.16$\pm$24.03*    & 66.24$\pm$24.08*           &  \\
				(i)  & \checkmark   & \checkmark  & \checkmark  & \textendash               		 &  & 69.94$\pm$21.37***    & 68.51$\pm$21.46**    & 69.71$\pm$21.42***  & 69.79$\pm$21.37***       &  \\
				(j)  & \checkmark   & \checkmark  & \checkmark  & \checkmark                &  & \textbf{73.24$\pm$21.35}  & \textbf{71.53$\pm$20.89} & \textbf{72.48$\pm$21.45}  & \textbf{73.21$\pm$21.13} &  \\ \cline{1-10}
			\end{tabular}
		}
		\begin{tablenotes}
			\footnotesize
			\item `\textendash' denotes which module is deleted from the proposed GAN model, and `$\checkmark$' denotes which module is remained. The asterisk in the \\ table indicate  significant difference between each pair of the two methods by paired t-tests (*$p < 0.05$, **$p < 0.01$, ***$p < 0.001$)
		\end{tablenotes}
	\end{threeparttable}
	\label{tab3}
\end{table*}

\section{Discussion}
\begin{figure*}[htbp]
	\center{\includegraphics[scale=0.18]{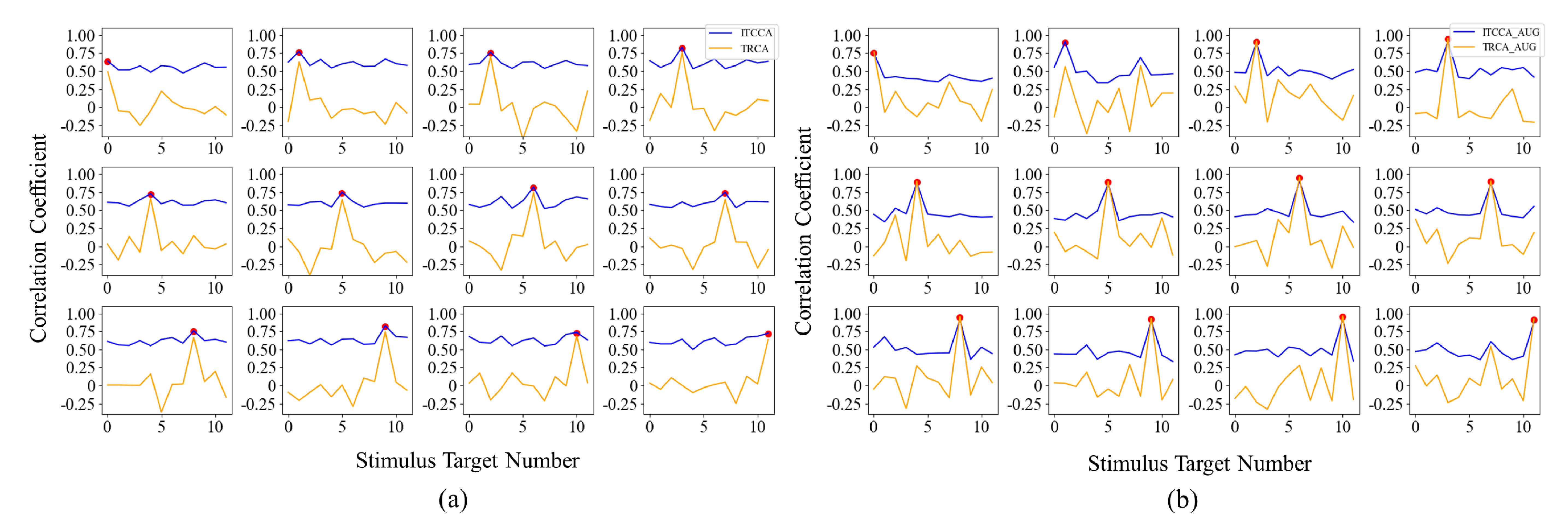}
		\caption{Correlation coefficients of a representative subject (subject 5 on small-scale dataset of Dial SSVEP Dataset) for two traditional classification methods as ITCCA and TRCA. (a) Original signals at 1.0 s, (b) Augmented signals at 2.0 s. The red dot marks the correlation coefficient at the frequency of gaze following target.}
		\label{fig7}}
\end{figure*}

\begin{figure*}[htbp]
	\center{\includegraphics[scale=0.12]{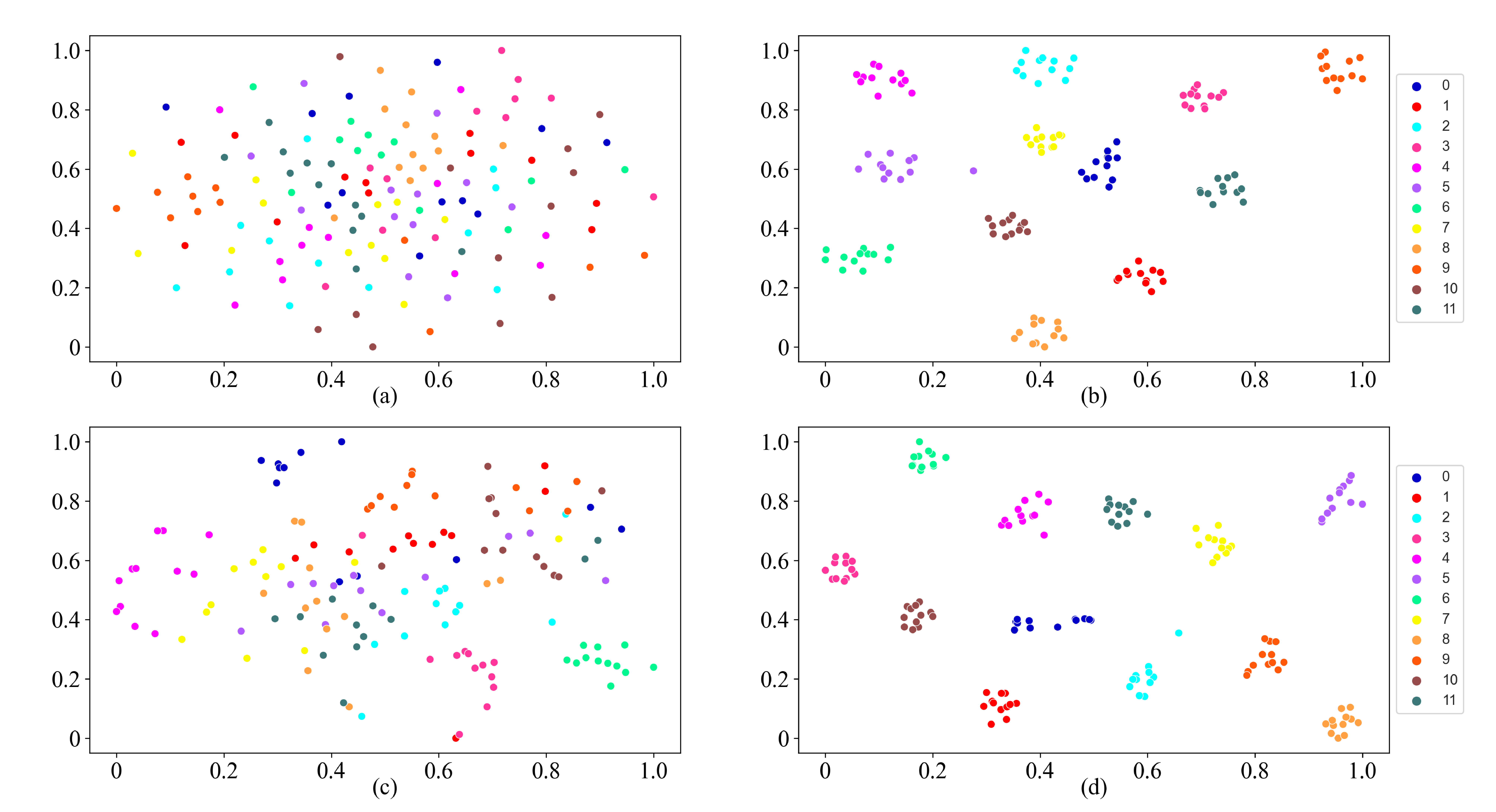}
		\caption{Feature distribution visualization of a representative subject (subject 5 on small-scale dataset of Dial SSVEP Dataset) using tSNE for EEGNet and C-CNN. (a) EEGNet using original signals at 1 s. (b) EEGNet using augmented signals at 2 s. (c) C-CNN using original signals at 1 s. (d) C-CNN using augmented signals at 2 s.}
		\label{fig8}}
\end{figure*}

In recent years, enhancing the classification performance of SSVEPs with limited calibration data is a hot research topic, which empower the practicality of various BCI applications \cite{pan2022SSVEPNet, wong2020msTRCA, wang2022SST, jorajuria2022OSTDA}. To this end, modelling an efficient classification method requiring less calibration data or create supplemented synthetic data to enlarge the size of the training dataset is the most commonly strategy. Nevertheless, in this study, we proposed a novel pipeline to address this issue. Specifically, we developed a GAN-based model, i.e., TEGAN, which could be used to extend the data length of SSVEP data. TEGAN seeks to learn the mapping relationship between short-length and long-length signals through an adversarial game training paradigm. Concretely, the generator is responsible for extracting signal features from the short-length SSVEP data and reconstructing the long-length SSVEP data. The discriminator assists the generator to improve the quality of the generated data by learning the discrepancy information between the real and fake long-time signal. To enhance the performance of TEGAN and mitigate the unstable training process of GAN, the two-stage training strategy and LeCam divergence regularization term are incorporated in the network design. We have designed several experiments to evaluate the effectiveness of TEGAN on two public datasets (4-class and 12-class), and achieved remarkable results under some certain conditions. However, we also have discovered some intriguing phenomena after analyzing the experimental results. To uncover the reasons behind these phenomena, here we discuss the potential value of our proposed method and analyze its current limitations, pointing out future research directions.

\subsection{Reducing the performance gap through reconstructed long-length SSVEP signals}
From Fig.~\ref{fig4} to Fig.~\ref{fig6}, we could find an interesting phenomenon that reconstructed long-length SSVEP signals are capable of reducing the classification performance gap among various classification methods. To explain this phenomenon, we visualize extracted features of a representative subject (subject 5 on Dial SSVEP Dataset) for four baseline classification methods. 

For traditional methods, we treat the correlation coefficient as extracted features. Fig.~\ref{fig7} illustrates the differences of correlation coefficient calculated by ITCCA and TRCA on the small-scale training dataset in the intra-subject experiments. Among them, Fig.~\ref{fig7} (a) and (b) present the visualization results of two methods using original signals at 1 s and using augmented signals at 2 s, respectively. It could be observed from Fig.~\ref{fig7} that there was a large gap in the correlation coefficients obtained by each category on the original signals, while the gap would be gradually narrowed on the augmented signals. Specifically, for the classification algorithms that previously had not obvious features of the target stimulus on the original signal, the features were further amplified on the augmented signals. For instance, when the ITCCA algorithm uses original signal at 1.0 s to classify the SSVEP data, the calculated correlation of stimulus target 10 is 0.74, the correlation coefficients of surrounding stimulus targets 0, 3, 6 and 9 are 0.68, 0.69, 0.66 and 0.71. The differences are less than 0.1. However, when the ITCCA algorithm uses augmented signal at 2.0 s to classify the SSVEP data stimulus target 10, the calculated correlation coefficient of stimulus target 10 is 0.92, while the correlation coefficients of target 0, 3, 6 and 9 are 0.43, 0.50, 0.51 and 0.42, the differences become much obvious. 

For DL-based methods, we utilize t-Stochastic Neighborhood Embedding (t-SNE) technique \cite{van2008tSNE} to visualize the extracted features in the penultimate layer. Fig.~\ref{fig8} exhibits the visualization results of feature distribution extracted by EEGNet and C-CNN on the small-scale training dataset in the intra-subject experiments. Specifically, Fig.~\ref{fig8} (a) and (c) show the visualization results of feature distribution using the original signal at 1.0 s, while the visualization results of feature distribution using the augmented signal at 2.0 s are displayed in Fig.~\ref{fig8} (b) and (d). It can be observed that, compared with the circumstances using the original signal, the feature distribution extracted by two DL methods using augmented signals has larger between-class distances and smaller within-class distances. Using the augmented signals as the inputs of DL methods leads to better clustering and class separation. Furthermore, the overlap between classes is extremely serious when the original signal is used as the network input, but there is no overlap between classes using the augmented signal. This indicates that the augmented signals, i.e. reconstructed long-length signals possess more discriminative features of the target stimulus compared to the original short-length signals. Consequently, it is easier for the classification algorithm to distinguish between target and non-target stimuli, which facilitates the improvement of the classification performance.

\begin{figure*}[htbp]
	\center{\includegraphics[scale=0.18]{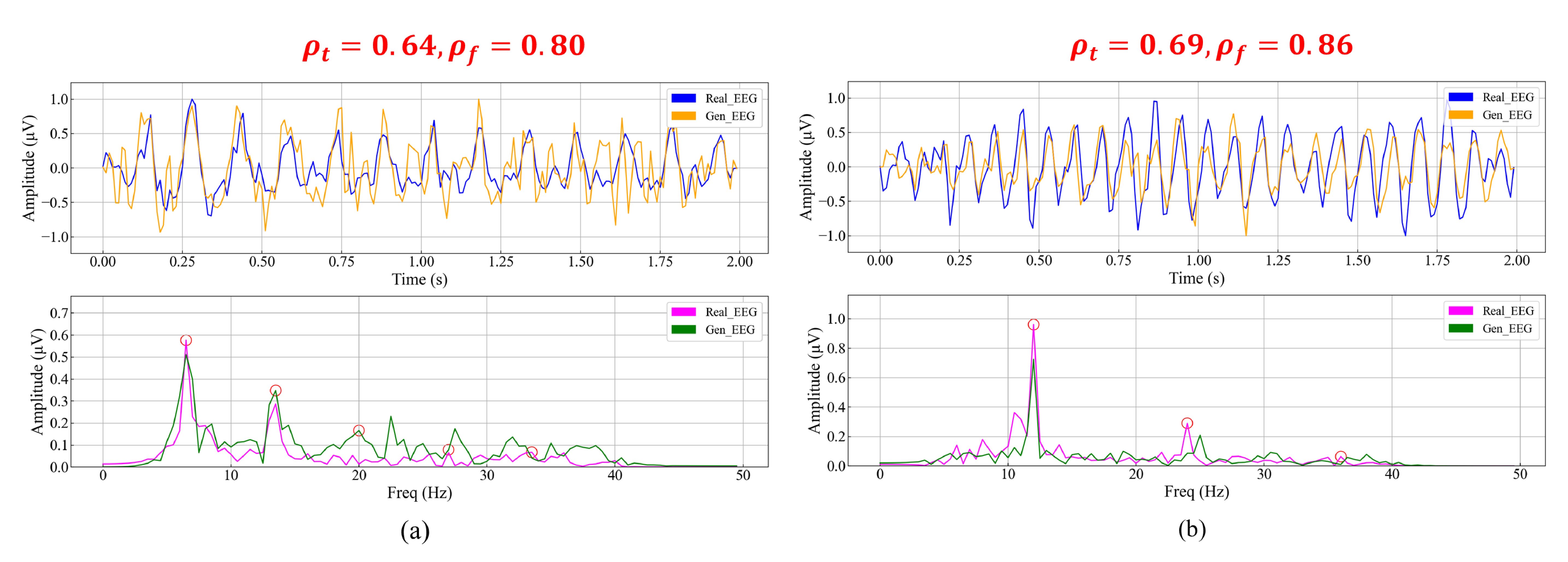}
		\caption{Comparison of time and frequency domain representation at Oz channel between real EEG data and generated EEG data on small-scale dataset of Direction SSVEP Dataset. (a) SSVEP signal at 6.67 Hz of subject 49, (b) SSVEP signal at 12.0 Hz of subject 10. The time window for the real EEG data is 2.0 s, while the generated data is transformed from the real EEG data at 1.0 s. In the frequency domain representation, red circles mark the fundamental frequency or harmonics. $\rho_t$ and $\rho_f$ denote the correlation coefficient of time and frequency domain representation between real long-length EEG and generated long-length EEG, respectively.}
		\label{fig9}}
\end{figure*}

\subsection{Investigating the quality of generated long-length SSVEP signals}
Generally, the amplitude of the fundamental frequency peak in the SSVEP signal is higher than that of the harmonic and subharmonic peaks \cite{zhang2022ARBCI}. However, based on the particularity that the harmonic component or subharmonic component of SSVEP signal is not obvious, we speculate that this feature would bring great trouble to GAN in the process of learning to generate SSVEP data. Deep neural networks may tend to preferentially capture the fundamental frequency information with dominated characteristics, while ignoring the harmonic and subharmonic information with less obvious characteristics. This is evident from the previous study \cite{aznan2019SSVEPGAN} that the SSVEP data generated by GAN are weaker than the original data in terms of the associated harmonics. To further verify this hypothesis in our study, we made a comparison of time and frequency domain representation between real SSVEP data and generated SSVEP data. To better visualize spectrum information, two SSVEP frequencies (6.67 Hz and 12.00 Hz) on Direction SSVEP Dataset were chosen for evaluation. Original signal at 2.0 s and augmented signal at 2.0 s (generated by 1.0 s original signal) were compared in our study, and the TEGAN was trained with the small-scale training dataset of the intra-subject experiments. All training trials at a single frequency are averaged and conduct Z-Score standardization for final visualization. As shown in Fig.~\ref{fig9}, we could observe that the original SSVEP signal and the generated SSVEP signal present high correlation both in the time domain and frequency domain (6.67 Hz: $\rho_t$=0.64, $\rho_f$=0.80; 12.0 Hz: $\rho_t$=0.64, $\rho_f$=0.86). However, both the original signal and the generated signal have significant peaks on the fundamental frequency in the frequency domain representation, while the identifiable harmonic components contained in the original signal are hard to reproduce in the generated signal, especially for those high-frequency harmonics. For example, the generated signal at 6.67 Hz has striking peaks on harmonics below 20 Hz, but there are no peaks at harmonic frequencies above 20 Hz and several unexpected peaks are emerged. This enlightens us that, in addition to ensuring that GAN pays attention to the fundamental frequency information, we can also improve the quality of the generated signal by emphasizing the harmonic response information in the generation process.

\subsection{Exploring the classification effect of extending the signal length multiple times}
The longer the signal length of SSVEPs, the weaker the interference from spontaneous EEG, which is part of the reason why training-free methods such as CCA and MSI algorithms could achieve better performance under long time-window lengths while losing efficacy with short time-window lengths \cite{lin2006CCA, zhang2014MSI}. For training-based methods, the issue of perturbation from spontaneous EEG in the short-length SSVEPs could be moderately alleviated via studying the subject-invariant or subject-variant properties \cite{bin2011ITCCA, zhang2011MCCA, nakanishi2017TRCA, zhang2018CORCA, waytowich2018EEGNet, ravi2020CCNN, guney2021tsDNN, pan2022SSVEPNet, chen2023SSVEPformer}. Moreover, compared to short-length SSVEPs, long-length SSVEPs have more distinct patterns and contain more information, which facilitates the feature extraction stage for training-based SSVEP classifiers. This leads to the phenomenon that while training-based methods can achieve more robust classification performance using short-length SSVEP, they will further achieve better classification performance on longer signal lengths.  In this study, we borrow this feature and aim to use TEGAN to learn robust and efficient information in short-length SSVEPs and reconstruct the corresponding extended long-length SSVEPs signals accordingly.  To maximize the utility of TEGAN as much as possible, we investigated the performance impact of using different extension factors ($F$=2, 3, 4) on various SSVEP classifiers. Based on the small-scale dataset of Dial SSVEP dataset, classification results with 0.5 short-length SSVEP, corresponding to its artificial and real long-length SSVEP at 1.0 s ($F$=2), 1.5 s ($F$=3), 2.0 s ($F$=4) are shown in Table~\ref{tab4}. When the extension factor increases from 2 to 4, the classification performance of the artificial long-length SSVEP shows a trend of first increasing and then decreasing, i.e., all classifiers achieve optimal performance when $F$=3. In addition, the classification performance gap between artificial long-length SSVEP and real long-length SSVEP signals is gradually widening. This suggests that although the performance of TEGAN significantly improves with increasing extension factors, learning the mapping relationship between short-length SSVEP and long-length SSVEP would also become more difficult. Seeking a more effective way to establish the mapping relationship to apply a larger extension factor to TEGAN would be a possible solution to fully tap the potential of TEGAN.

\begin{table*}[]
	\caption{\centering{Classification results of different baseline classifiers using various extension factors ($F$=2, 3, 4) on small-scale dataset of Dial SSVEP dataset. The original signal length was set to 0.5 s, corresponding to its augmented and original signal length at 1.0 s ($F$=2), 1.5 s ($F$=3), 2.0 s ($F$=4), respectively. ORG and AUG is the result of using real long-length EEG and generated long-length EEG, respectively. The highest accuracy (in bold) indicates the optimal scheme for this condition.}}
	\renewcommand{\arraystretch}{1.5}
	\scalebox{0.8}{
	\begin{tabular}{c|c|cc|cc|cc}
		\hline
		\multirow{2}{*}{Method}    &             & \multicolumn{2}{c|}{$F$=2}  & \multicolumn{2}{c|}{$F$=3}  & \multicolumn{2}{c}{$F$=4}   \\ \cline{2-8} 
		& 0.5 s & 1.0 s (AUG) & 1.0 s (ORG) & 1.5 s (AUG) & 1.5 s (ORG) & 2.0 s (AUG) & 2.0 s (ORG) \\ \hline
		ITCCA                      & 42.44 $\pm$ 18.77 & \textbf{73.24 $\pm$ 21.35} & 67.01 $\pm$ 24.66 & 74.74 $\pm$ 21.15 & \textbf{80.31 $\pm$ 20.99} & 73.12 $\pm$ 21.48 & \textbf{85.08 $\pm$ 18.23} \\ \hline
		TRCA                       & 64.99 $\pm$ 21.90 & 71.53 $\pm$ 20.89 & \textbf{82.93 $\pm$ 17.39} & 73.82 $\pm$ 21.18 & \textbf{90.00 $\pm$ 12.89} & 71.40 $\pm$ 20.59 & \textbf{92.47 $\pm$ 10.37} \\ \hline
		EEGNet                     & 35.44 $\pm$ 13.93 & \textbf{72.48 $\pm$ 21.45} & 55.63 $\pm$ 23.58 & \textbf{73.60 $\pm$ 21.22} & 65.58 $\pm$ 25.18 & \textbf{72.35 $\pm$ 21.21} & 71.42 $\pm$ 23.91 \\ \hline
		\multicolumn{1}{l|}{C-CNN} & 53.26 $\pm$ 19.65 & \textbf{73.21 $\pm$ 21.13} & 67.51 $\pm$ 21.69 & 74.50 $\pm$ 21.35 & \textbf{77.63 $\pm$ 21.75} & 73.44 $\pm$ 20.74 & \textbf{81.05 $\pm$ 21.65} \\ \hline
	\end{tabular}
	}
\label{tab4}
\end{table*}

\subsection{Limitation and future work}
According to the limitations and challenges of current study, the following aspects could be further improved in the future. Firstly, to the best of our knowledge, this is the first research using GAN technology to generate more than 10 categories of SSVEP data (Dial SSVEP dataset). However, with the increasing number of categories, it would become extremely difficult for GAN to simulate the real data distribution. Therefore, we should strive to generate SSVEP data with more categories, such as Benchmark \cite{wang2016Benchmark} and BETA dataset \cite{liu2020BETA}. Secondly, filter bank technology should be incorporated in the framework design as well, which could help the GAN model enhance the harmonic information that has an important contribution to the identification process, thus improving the quality of generated data. Furthermore, the most advanced transfer learning technique, such as improved two-stage training strategy \cite{guney2023TLDNN}, could be utilized to build a high-performance BCI system with shorter calibration time. Besides, building a plug-and-play BCI system that achieves zero-calibration for new users is always the ultimate goal for the BCI researchers. Hence, an effective scheme for applying our augmentation method in generating cross-subject data should be proposed. Last but not least, since the TEGAN model could reduce the performance gap of various frequency recognition algorithms, the simplest algorithms could be used to achieve the recognition performance comparable with complex algorithms in the actual application deployment process to cut down the budget of development SSVEP-BCI systems.

\section{Conclusion}
In this study, we proposed a novel GAN-based augmentation model to enhance the performance of SSVEP-based BCI. Rather than enlarging the amount of training dataset, the proposed augmentation method served as a signal converter that could be employed to extend data length. We incorporated a novel U-Net generator architecture and auxiliary classifier into the network architecture, and introduced a two-stage training strategy and the LeCam-divergence regularization term during the network implementation. Extensive experimental results based on a 4-class and a 12-class SSVEP datasets demonstrate that the proposed augmentation method could significantly improve the recognition accuracy of traditional methods and deep learning methods with limited subject-specific data, and could significantly shorten the calibration time. Furthermore, the proposed method holds promise to reduce the classification performance gap of various frequency recognition methods. This study could shed light on future research to boost the performance of the SSVEP-based BCI by extending EEG data length by generative models as GANs. 

\printcredits

\section*{Acknowledgement}
This work was supported in part by the National Natural Science Foundation of China under Grant No.62076209.

\bibliographystyle{unsrt}

\bibliography{reference}

\begin{thebibliography}{10}

\bibitem{wolpaw2000BCI}
Jonathan~R Wolpaw, Niels Birbaumer, William~J Heetderks, Dennis~J McFarland,
  P~Hunter Peckham, Gerwin Schalk, Emanuel Donchin, Louis~A Quatrano, Charles~J
  Robinson, Theresa~M Vaughan, et~al.
\newblock Brain-computer interface technology: a review of the first
  international meeting.
\newblock {\em IEEE transactions on rehabilitation engineering}, 8(2):164--173,
  2000.

\bibitem{hwang2013EEG}
Han-Jeong Hwang, Soyoun Kim, Soobeom Choi, and Chang-Hwan Im.
\newblock {EEG}-based brain-computer interfaces: a thorough literature survey.
\newblock {\em International Journal of Human-Computer Interaction},
  29(12):814--826, 2013.

\bibitem{abiri2019EEG}
Reza Abiri, Soheil Borhani, Eric~W Sellers, Yang Jiang, and Xiaopeng Zhao.
\newblock A comprehensive review of {EEG}-based brain--computer interface
  paradigms.
\newblock {\em Journal of neural engineering}, 16(1):011001, 2019.

\bibitem{ang2008FBCSP}
Kai~Keng Ang, Zheng~Yang Chin, Haihong Zhang, and Cuntai Guan.
\newblock Filter bank common spatial pattern ({FBCSP}) in brain-computer
  interface.
\newblock In {\em 2008 IEEE international joint conference on neural networks
  (IEEE world congress on computational intelligence)}, pages 2390--2397, Hong
  Kong, China, 01--08 June 2008. IEEE.

\bibitem{nijboer2008P300}
Femke Nijboer, EW~Sellers, J{\"u}rgen Mellinger, Mary~Ann Jordan, Tamara Matuz,
  Adrian Furdea, Sebastian Halder, Ursula Mochty, DJ~Krusienski, TM~Vaughan,
  et~al.
\newblock A {P300}-based brain--computer interface for people with amyotrophic
  lateral sclerosis.
\newblock {\em Clinical neurophysiology}, 119(8):1909--1916, 2008.

\bibitem{kim2011ASSR}
Do-Won Kim, Han-Jeong Hwang, Jeong-Hwan Lim, Yong-Ho Lee, Ki-Young Jung, and
  Chang-Hwan Im.
\newblock Classification of selective attention to auditory stimuli: toward
  vision-free brain--computer interfacing.
\newblock {\em Journal of neuroscience methods}, 197(1):180--185, 2011.

\bibitem{zhang2012MFSC}
Yangsong Zhang, Peng Xu, Tiejun Liu, Jun Hu, Rui Zhang, and Dezhong Yao.
\newblock Multiple frequencies sequential coding for ssvep-based brain-computer
  interface.
\newblock {\em PloS one}, 7(3):e29519, 2012.

\bibitem{lim2013SSVEP}
Jeong-Hwan Lim, Han-Jeong Hwang, Chang-Hee Han, Ki-Young Jung, and Chang-Hwan
  Im.
\newblock Classification of binary intentions for individuals with impaired
  oculomotor function:‘eyes-closed’{SSVEP}-based brain--computer interface
  ({BCI}).
\newblock {\em Journal of neural engineering}, 10(2):026021, 2013.

\bibitem{kwak2017CNN}
No-Sang Kwak, Klaus-Robert M{\"u}ller, and Seong-Whan Lee.
\newblock A convolutional neural network for steady state visual evoked
  potential classification under ambulatory environment.
\newblock {\em PloS one}, 12(2):e0172578, 2017.

\bibitem{dang2021MHLCNN}
Weidong Dang, Mengyu Li, Dongmei Lv, Xinlin Sun, and Zhongke Gao.
\newblock Mhlcnn: {Multi}-harmonic linkage {CNN} model for {SSVEP} and {SSMVEP}
  signal classification.
\newblock {\em IEEE Transactions on Circuits and Systems II: Express Briefs},
  69(1):244--248, 2021.

\bibitem{nakanishi2015eCCA}
Masaki Nakanishi, Yijun Wang, Yu-Te Wang, and Tzyy-Ping Jung.
\newblock A comparison study of canonical correlation analysis based methods
  for detecting steady-state visual evoked potentials.
\newblock {\em PloS one}, 10(10):e0140703, 2015.

\bibitem{chen2015Speller}
Xiaogang Chen, Yijun Wang, Masaki Nakanishi, Xiaorong Gao, Tzyy-Ping Jung, and
  Shangkai Gao.
\newblock High-speed spelling with a noninvasive brain--computer interface.
\newblock {\em Proceedings of the national academy of sciences},
  112(44):E6058--E6067, 2015.

\bibitem{kim2019SmartHome}
Minju Kim, Min-Ki Kim, Minho Hwang, Hyun-Young Kim, Jeongho Cho, and Sung-Phil
  Kim.
\newblock Online home appliance control using {EEG}-based brain--computer
  interfaces.
\newblock {\em Electronics}, 8(10):1101, 2019.

\bibitem{lalor2005GAME}
Edmund~C Lalor, Simon~P Kelly, Ciar{\'a}n Finucane, Robert Burke, Ray Smith,
  Richard~B Reilly, and Gary Mcdarby.
\newblock Steady-state vep-based brain-computer interface control in an
  immersive {3D} gaming environment.
\newblock {\em EURASIP Journal on Advances in Signal Processing},
  2005(19):1--9, 2005.

\bibitem{Pan2023Survey}
Yudong Pan, Jianbo Chen, and Yangsong Zhang.
\newblock A {Survey} of deep learning-based classification methods for
  steady-state visual evoked potentials.
\newblock {\em Brain-Apparatus Communication: A Journal of Bacomics},
  2(1):2181102, 2023.

\bibitem{hakvoort2011PSDA}
Gido Hakvoort, Boris Reuderink, and Michel Obbink.
\newblock {Comparison of PSDA and CCA detection methods in a SSVEP-based
  BCI-system}.
\newblock {\em Centre for Telematics \& Information Technology University of
  Twente}, 2011.

\bibitem{lin2006CCA}
Zhonglin Lin, Changshui Zhang, Wei Wu, and Xiaorong Gao.
\newblock Frequency recognition based on canonical correlation analysis for
  {SSVEP}-based {BCIs}.
\newblock {\em IEEE transactions on biomedical engineering}, 53(12):2610--2614,
  2006.

\bibitem{zhang2014MSI}
Yangsong Zhang, Peng Xu, Kaiwen Cheng, and Dezhong Yao.
\newblock Multivariate synchronization index for frequency recognition of
  {SSVEP}-based brain--computer interface.
\newblock {\em Journal of neuroscience methods}, 221:32--40, 2014.

\bibitem{chen2021implementing}
Yonghao Chen, Chen Yang, Xiaochen Ye, Xiaogang Chen, Yijun Wang, and Xiaorong
  Gao.
\newblock Implementing a calibration-free {SSVEP}-based {BCI} system with 160
  targets.
\newblock {\em Journal of Neural Engineering}, 18(4):046094, 2021.

\bibitem{bin2011ITCCA}
Guangyu Bin, Xiaorong Gao, Yijun Wang, Yun Li, Bo~Hong, and Shangkai Gao.
\newblock A high-speed {BCI} based on code modulation {VEP}.
\newblock {\em Journal of neural engineering}, 8(2):025015, 2011.

\bibitem{nakanishi2017TRCA}
Masaki Nakanishi, Yijun Wang, Xiaogang Chen, Yu-Te Wang, Xiaorong Gao, and
  Tzyy-Ping Jung.
\newblock Enhancing detection of {SSVEPs} for a high-speed brain speller using
  task-related component analysis.
\newblock {\em IEEE Transactions on Biomedical Engineering}, 65(1):104--112,
  2017.

\bibitem{zhang2018CORCA}
Yangsong Zhang, Daqing Guo, Fali Li, Erwei Yin, Yu~Zhang, Peiyang Li, Qibin
  Zhao, Toshihisa Tanaka, Dezhong Yao, and Peng Xu.
\newblock Correlated component analysis for enhancing the performance of
  {SSVEP}-based brain-computer interface.
\newblock {\em IEEE Transactions on Neural Systems and Rehabilitation
  Engineering}, 26(5):948--956, 2018.

\bibitem{chiang2021LST}
Kuan-Jung Chiang, Chun-Shu Wei, Masaki Nakanishi, and Tzyy-Ping Jung.
\newblock Boosting template-based {SSVEP} decoding by cross-domain transfer
  learning.
\newblock {\em Journal of Neural Engineering}, 18(1):016002, 2021.

\bibitem{wong2020stCCA}
Chi~Man Wong, Ze~Wang, Boyu Wang, Ka~Fai Lao, Agostinho Rosa, Peng Xu,
  Tzyy-Ping Jung, CL~Philip Chen, and Feng Wan.
\newblock Inter-and intra-subject transfer reduces calibration effort for
  high-speed {SSVEP}-based {BCIs}.
\newblock {\em IEEE Transactions on Neural Systems and Rehabilitation
  Engineering}, 28(10):2123--2135, 2020.

\bibitem{wong2021tlCCA}
Chi~Man Wong, Ze~Wang, Agostinho~C Rosa, CL~Philip Chen, Tzyy-Ping Jung, Yong
  Hu, and Feng Wan.
\newblock Transferring subject-specific knowledge across stimulus frequencies
  in {SSVEP}-based {BCIs}.
\newblock {\em IEEE Transactions on Automation Science and Engineering},
  18(2):552--563, 2021.

\bibitem{yan2022CSSFT}
Wenqiang Yan, Yongcheng Wu, Chenghang Du, and Guanghua Xu.
\newblock Cross-subject spatial filter transfer method for {SSVEP}-{EEG}
  feature recognition.
\newblock {\em Journal of Neural Engineering}, 19(3):036008, 2022.

\bibitem{waytowich2018EEGNet}
Nicholas Waytowich, Vernon~J Lawhern, Javier~O Garcia, Jennifer Cummings, Josef
  Faller, Paul Sajda, and Jean~M Vettel.
\newblock Compact convolutional neural networks for classification of
  asynchronous steady-state visual evoked potentials.
\newblock {\em Journal of neural engineering}, 15(6):066031, 2018.

\bibitem{ravi2020CCNN}
Aravind Ravi, Nargess~Heydari Beni, Jacob Manuel, and Ning Jiang.
\newblock Comparing user-dependent and user-independent training of {CNN} for
  {SSVEP} {BCI}.
\newblock {\em Journal of neural engineering}, 17(2):026028, 2020.

\bibitem{chen2023SSVEPformer}
Jianbo Chen, Yangsong Zhang, Yudong Pan, Peng Xu, and Cuntai Guan.
\newblock A transformer-based deep neural network model for ssvep
  classification.
\newblock {\em Neural Networks}, 164:521--534, 2023.

\bibitem{pan2022SSVEPNet}
Yudong Pan, Jianbo Chen, Yangsong Zhang, and Yu~Zhang.
\newblock An efficient {CNN}-{LSTM} network with spectral normalization and
  label smoothing technologies for {SSVEP} frequency recognition.
\newblock {\em Journal of Neural Engineering}, 19(5):056014, 2022.

\bibitem{luo2022SAME}
Ruixin Luo, Minpeng Xu, Xiaoyu Zhou, Xiaolin Xiao, Tzyy-Ping Jung, and Dong
  Ming.
\newblock Data augmentation of {SSVEPs} using source aliasing matrix estimation
  for brain-computer interfaces.
\newblock {\em IEEE Transactions on Biomedical Engineering}, pages 1--10, 2022.

\bibitem{thirumuruganathan2020DGM}
Saravanan Thirumuruganathan, Shohedul Hasan, Nick Koudas, and Gautam Das.
\newblock Approximate query processing for data exploration using deep
  generative models.
\newblock In {\em 2020 IEEE 36th international conference on data engineering
  (ICDE)}, pages 1309--1320, Dallas, TX, USA, 20--24 April 2020. IEEE.

\bibitem{salimans2017PixelCNN}
Tim Salimans, Andrej Karpathy, Xi~Chen, and Diederik~P Kingma.
\newblock Pixelcnn++: {Improving} the pixelcnn with discretized logistic
  mixture likelihood and other modifications.
\newblock {\em arXiv preprint arXiv:1701.05517}, 2017.

\bibitem{kingma2013VAE}
Diederik~P Kingma and Max Welling.
\newblock Auto-encoding variational bayes.
\newblock {\em arXiv preprint arXiv:1312.6114}, 2013.

\bibitem{goodfellow2014GAN}
Ian~J Goodfellow, Jean Pouget-Abadie, Mehdi Mirza, Bing Xu, David Warde-Farley,
  Sherjil Ozair, Aaron~C Courville, and Yoshua Bengio.
\newblock {Generative adversarial nets}.
\newblock {\em Advances in neural information processing systems},
  27:2672--2680, 2014.

\bibitem{ho2020DDPM}
Jonathan Ho, Ajay Jain, and Pieter Abbeel.
\newblock Denoising diffusion probabilistic models.
\newblock {\em Advances in Neural Information Processing Systems},
  33:6840--6851, 2020.

\bibitem{hartmann2018EEGGAN}
Kay~Gregor Hartmann, Robin~Tibor Schirrmeister, and Tonio Ball.
\newblock {EEG-GAN}: Generative adversarial networks for electroencephalograhic
  (eeg) brain signals.
\newblock {\em arXiv preprint arXiv:1806.01875}, 2018.

\bibitem{aznan2019SSVEPGAN}
Nik Khadijah~Nik Aznan, Amir Atapour-Abarghouei, Stephen Bonner, Jason~D
  Connolly, Noura Al~Moubayed, and Toby~P Breckon.
\newblock Simulating brain signals: {Creating} synthetic eeg data via
  neural-based generative models for improved ssvep classification.
\newblock In {\em 2019 International Joint Conference on Neural Networks
  (IJCNN)}, pages 1--8, Budapest, Hungary, 14--19 July 2019. IEEE.

\bibitem{aznan2021SISGAN}
Nik Khadijah~Nik Aznan, Amir Atapour-Abarghouei, Stephen Bonner, Jason~D
  Connolly, and Toby~P Breckon.
\newblock {Leveraging Synthetic Subject Invariant EEG Signals for Zero
  Calibration BCI}.
\newblock In {\em 2020 25th International Conference on Pattern Recognition
  (ICPR)}, pages 10418--10425, Milan, Italy, 10--15 Januray 2021. IEEE.

\bibitem{kwon2022StarGAN}
Jinuk Kwon and Chang-Hwan Im.
\newblock {Novel Signal-to-Signal Translation Method Based on StarGAN to
  Generate Artificial EEG for SSVEP-Based Brain-Computer Interfaces}.
\newblock {\em Expert Systems with Applications}, 203:117574, 2022.

\bibitem{guney2021tsDNN}
Osman~Berke Guney, Muhtasham Oblokulov, and Huseyin Ozkan.
\newblock A deep neural network for ssvep-based brain-computer interfaces.
\newblock {\em IEEE Transactions on Biomedical Engineering}, 69(2):932--944,
  2021.

\bibitem{lee2019GIGADataset}
Min-Ho Lee, O-Yeon Kwon, Yong-Jeong Kim, Hong-Kyung Kim, Young-Eun Lee, John
  Williamson, Siamac Fazli, and Seong-Whan Lee.
\newblock {EEG dataset and OpenBMI toolbox for three BCI paradigms: An
  investigation into BCI illiteracy}.
\newblock {\em GigaScience}, 8(5):giz002, 2019.

\bibitem{tanaka2013TRCA}
Hirokazu Tanaka, Takusige Katura, and Hiroki Sato.
\newblock Task-related component analysis for functional neuroimaging and
  application to near-infrared spectroscopy data.
\newblock {\em NeuroImage}, 64:308--327, 2013.

\bibitem{lawhern2018EEGNet}
Vernon~J Lawhern, Amelia~J Solon, Nicholas~R Waytowich, Stephen~M Gordon,
  Chou~P Hung, and Brent~J Lance.
\newblock {EEGNet}: a compact convolutional neural network for {EEG}-based
  brain--computer interfaces.
\newblock {\em Journal of neural engineering}, 15(5):056013, 2018.

\bibitem{zhang20213DCNN}
Yangsong Zhang, Huan Cai, Li~Nie, Peng Xu, Sirui Zhao, and Cuntai Guan.
\newblock An end-to-end {3D} convolutional neural network for decoding
  attentive mental state.
\newblock {\em Neural Networks}, 144:129--137, 2021.

\bibitem{odena2017ACGAN}
Augustus Odena, Christopher Olah, and Jonathon Shlens.
\newblock Conditional image synthesis with auxiliary classifier gans.
\newblock In {\em International conference on machine learning}, volume~70,
  pages 2642--2651, Sydney, Australia, 06--11 August 2017. PMLR.

\bibitem{miyato2018SN}
Takeru Miyato, Toshiki Kataoka, Masanori Koyama, and Yuichi Yoshida.
\newblock {Spectral normalization for generative adversarial networks}.
\newblock {\em arXiv preprint arXiv:1802.05957}, 2018.

\bibitem{KaiMing2015PReLU}
Kaiming He, Xiangyu Zhang, Shaoqing Ren, and Jian Sun.
\newblock {Delving Deep into Rectifiers: Surpassing Human-Level Performance on
  ImageNet Classification}.
\newblock In {\em Proceedings of the IEEE International Conference on Computer
  Vision (ICCV)}, Santiago, Chile, 11--18 December 2015. IEEE.

\bibitem{dumoulin2016CBN}
Vincent Dumoulin, Jonathon Shlens, and Manjunath Kudlur.
\newblock A learned representation for artistic style.
\newblock {\em arXiv preprint arXiv:1610.07629}, 2016.

\bibitem{ronneberger2015UNet}
Olaf Ronneberger, Philipp Fischer, and Thomas Brox.
\newblock U-net: {Convolutional} networks for biomedical image segmentation.
\newblock In {\em International Conference on Medical image computing and
  computer-assisted intervention}, pages 234--241, Munich, Germany, 05-09
  October 2015. Springer.

\bibitem{tseng2021LeCam}
Hung-Yu Tseng, Lu~Jiang, Ce~Liu, Ming-Hsuan Yang, and Weilong Yang.
\newblock Regularizing generative adversarial networks under limited data.
\newblock In {\em Proceedings of the IEEE/CVF Conference on Computer Vision and
  Pattern Recognition}, pages 7921--7931, June 2021.

\bibitem{wong2020msTRCA}
Chi~Man Wong, Feng Wan, Boyu Wang, Ze~Wang, Wenya Nan, Ka~Fai Lao, Peng~Un Mak,
  Mang~I Vai, and Agostinho Rosa.
\newblock Learning across multi-stimulus enhances target recognition methods in
  {SSVEP}-based {BCIs}.
\newblock {\em Journal of neural engineering}, 17(1):016026, 2020.

\bibitem{wang2022SST}
Ze~Wang, Chi~Man Wong, Agostinho Rosa, Tao Qian, Tzyy-Ping Jung, and Feng Wan.
\newblock {Stimulus-Stimulus Transfer Based on Time-Frequency-Joint
  Representation in SSVEP-Based BCIs}.
\newblock {\em IEEE Transactions on Biomedical Engineering}, (2):603--615,
  2022.

\bibitem{jorajuria2022OSTDA}
Tania Jorajur{\'\i}a, Mina~Jamshidi Idaji, Zafer {\.I}{\c{s}}can, Marisol
  G{\'o}mez, Vadim~V Nikulin, and Carmen Vidaurre.
\newblock {Oscillatory Source Tensor Discriminant Analysis (OSTDA): A
  regularized tensor pipeline for SSVEP-based BCI systems}.
\newblock {\em Neurocomputing}, 492:664--675, 2022.

\bibitem{van2008tSNE}
Laurens Van~der Maaten and Geoffrey Hinton.
\newblock {Visualizing data using t-SNE}.
\newblock {\em Journal of machine learning research}, 9(11), 2008.

\bibitem{zhang2022ARBCI}
Rui Zhang, Zongxin Xu, Lipeng Zhang, Lijun Cao, Yuxia Hu, Beihan Lu, Li~Shi,
  Dezhong Yao, and Xincan Zhao.
\newblock The effect of stimulus number on the recognition accuracy and
  information transfer rate of {SSVEP}--{BCI} in augmented reality.
\newblock {\em Journal of Neural Engineering}, 19(3):036010, 2022.

\bibitem{zhang2011MCCA}
Yu~Zhang, Guoxu Zhou, Qibin Zhao, Akinari Onishi, Jing Jin, Xingyu Wang, and
  Andrzej Cichocki.
\newblock Multiway canonical correlation analysis for frequency components
  recognition in {SSVEP}-based {BCIs}.
\newblock In {\em International Conference on Neural information processing},
  pages 287--295, Shanghai, China, 13--17 November 2011. Springer.

\bibitem{wang2016Benchmark}
Yijun Wang, Xiaogang Chen, Xiaorong Gao, and Shangkai Gao.
\newblock A benchmark dataset for {SSVEP}-based brain--computer interfaces.
\newblock {\em IEEE Transactions on Neural Systems and Rehabilitation
  Engineering}, 25(10):1746--1752, 2016.

\bibitem{liu2020BETA}
Bingchuan Liu, Xiaoshan Huang, Yijun Wang, Xiaogang Chen, and Xiaorong Gao.
\newblock {BETA}: A large benchmark database toward {SSVEP}-{BCI} application.
\newblock {\em Frontiers in neuroscience}, 14:627, 2020.

\bibitem{guney2023TLDNN}
Osman~Berke Guney and Huseyin Ozkan.
\newblock {Transfer learning of an ensemble of DNNs for SSVEP BCI spellers
  without user-specific training}.
\newblock {\em Journal of Neural Engineering}, 20(1):016013, 2023.

\end{thebibliography}

\end{sloppypar}
\end{document}